\documentclass[journal=jacsat,manuscript=article]{achemso}

\usepackage[version=3]{mhchem} 
\usepackage[noend]{algpseudocode}
\usepackage{algorithm}
\usepackage{algpseudocode}

\usepackage{float}
\usepackage{placeins}
\usepackage{color}
\usepackage[usenames,dvipsnames,svgnames,table]{xcolor}
\usepackage{colortbl}
\usepackage{bm}
\usepackage{bbm}
\usepackage{xspace}
\usepackage{amsmath}
\usepackage{mathtools}
\usepackage{amssymb}
\usepackage{amsbsy}
\usepackage{eucal}
\usepackage{mathrsfs}
\usepackage{tcolorbox}
\usepackage{wrapfig}
\usepackage{pdfpages}
\usepackage{graphicx}
\usepackage{caption}
\usepackage{changepage}
\usepackage{multirow}
\usepackage{lipsum}
\usepackage{cancel}
\usepackage{subcaption}

\usepackage{stmaryrd}
\usepackage{grffile}

\usepackage{hyperref}
\usepackage{pgf}
\usepackage{pgfgantt}

\usepackage{verbatim}
\usepackage{siunitx}

\usepackage{listings}

\usepackage[normalem]{ulem}

\usepackage{longtable}
\usepackage{caption}

\definecolor{cadetblue}{RGB}{47,85,151}
\definecolor{lightblue}{RGB}{180,199,231}

\newlength{\bibitemsep}
\newlength{\bibparskip}
\let\oldthebibliography\thebibliography
\renewcommand\thebibliography[1]{%
  \oldthebibliography{#1}%
  \setlength{\parskip}{\bibitemsep}%
  \setlength{\itemsep}{\bibparskip}%
}

\def\1{\ensuremath{_1}}
\def\2{\ensuremath{_2}}
\def\3{\ensuremath{_3}}
\def\4{\ensuremath{_4}}
\def\5{\ensuremath{_5}}
\def\6{\ensuremath{_6}}
\def\7{\ensuremath{_7}}
\def\8{\ensuremath{_8}}

\newcommand{\C}{{\ensuremath{\mathrm{C}}\xspace}}
\renewcommand{\H}{{\ensuremath{\mathrm{H}}\xspace}}
\renewcommand{\O}{{\ensuremath{\mathrm{O}}\xspace}}

\newcommand{\bn}{\begin{enumerate}}
\newcommand{\en}{\end{enumerate}}
\newcommand{\bi}{\begin{itemize}}
\newcommand{\ei}{\end{itemize}}
\newcommand{\bdi}{\begin{dashItemize}}
\newcommand{\edi}{\end{dashItemize}}

\newcommand{\gae}{\lower 2pt \hbox{$\, \buildrel {\scriptstyle >}\over {\scriptstyle \sim}\,$}}
\newcommand{\lae}{\lower 2pt \hbox{$\, \buildrel {\scriptstyle <}\over {\scriptstyle \sim}\,$}}

\newcommand{\be}{\begin{equation}}
\newcommand{\ee}{\end{equation}}
\newcommand{\benn}{\begin{equation*}}           
\newcommand{\eenn}{\end{equation*}}
\newcommand{\bea}{\begin{eqnarray}}
\newcommand{\eea}{\end{eqnarray}}
\newcommand{\bean}{\begin{eqnarray*}}
\newcommand{\eean}{\end{eqnarray*}}

\newcommand{\invisible}[1]{ }

\newtcbox{\mybox}{
 nobeforeafter,colframe=black,colback=white,boxrule=0.5pt,arc=0pt,
 boxsep=3pt,left=1pt,right=1pt,top=0pt,bottom=0pt,tcbox raise base
}

\tikzstyle{startstop} = [rectangle, rounded corners, minimum width=5cm, minimum height=1.5cm,text centered, draw=black, fill=red!30]
\tikzstyle{io} = [trapezium, trapezium left angle=70, trapezium right angle=110, minimum width=3cm, minimum height=1cm, text centered, text width=2cm, draw=black, fill=blue!30]
\tikzstyle{process} = [rectangle, minimum width=3cm, minimum height=1cm, text centered, text width=2cm, draw=black, fill=orange!30]
\tikzstyle{decision} = [diamond, minimum width=1cm, minimum height=1cm, text centered, text width=2cm, draw=black, fill=green!30]
\tikzstyle{arrow} = [thick,->,>=stealth]
\tikzstyle{line} = [thick,-,>=stealth]
\tikzstyle{dashed} = [thick,..,>=stealth]

\graphicspath{{./fig/}}

\newcommand{\RR}{\mathbb{R}}

\makeatletter
\newcommand*\bigcdot{\mathpalette\bigcdot@{.5}}
\newcommand*\bigcdot@[2]{\mathbin{\vcenter{\hbox{\scalebox{#2}{$\m@th#1\bullet$}}}}}
\makeatother

\usepackage[sort&compress]{natbib}

\author{Christian Devereux}
\author{Yoona Yang}
\author{Carles Mart\'i}
\author{Judit Z\'ador}
\affiliation{Combustion Research Facility, Sandia National Laboratories, Livermore, CA 94551, USA}
\author{Michael S. Eldred}
\affiliation{Sandia National Laboratories, Albuquerque, NM 87185, USA}
\author{Habib N. Najm}
\email{hnnajm@sandia.gov}
\affiliation{Sandia National Laboratories, Livermore, CA 94551, USA}

\title{Force Training Neural Network Potential Energy Surface Models}

\begin{document}

\begin{tocentry}
\scriptsize

\includegraphics[width=\textwidth]{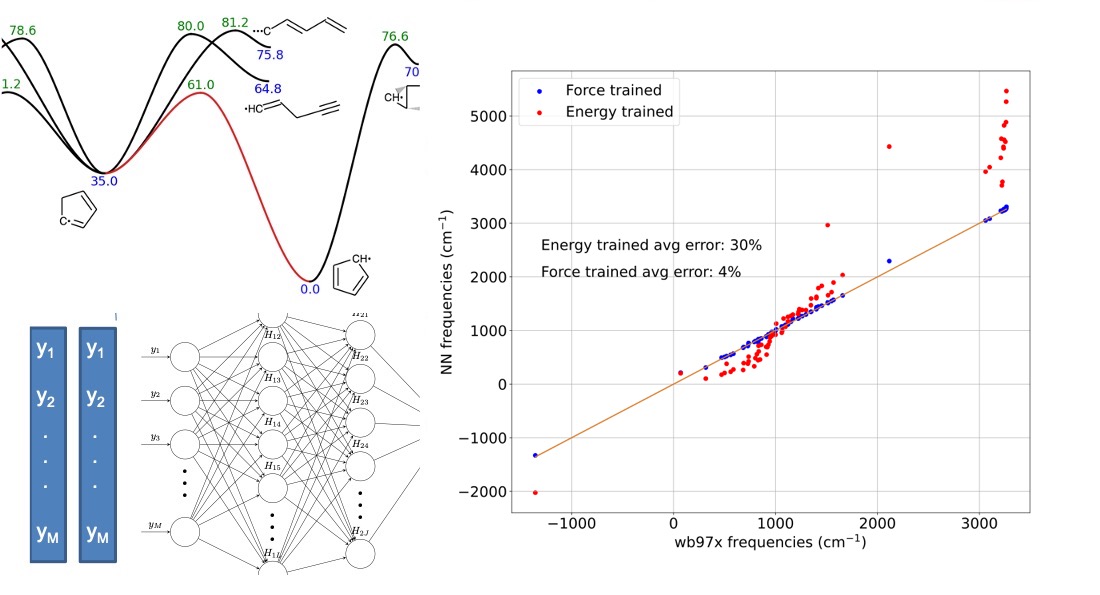}

\end{tocentry}

\begin{abstract}
Machine learned chemical potentials have shown great promise as alternatives to conventional computational chemistry methods to represent the potential energy of a given atomic or molecular system as a function of its geometry.  
However, such potentials are only as good as the data they are trained on, and building a comprehensive training set can be a costly process. Therefore, it is important to extract as much information from training data as possible without further increasing the computational cost. One way to accomplish this is by training on molecular forces in addition to energies.  This allows for three additional labels per atom within the molecule. Here we develop a neural network potential energy surface for studying a hydrogen transfer reaction between two conformers of $\C_5\H_5$. We show that, for a much smaller training set, force training can greatly improve the accuracy of the model compared to only training to energies. We also demonstrate the importance of choosing the proper force to energy weight ratio for the loss function to minimize the model test error. 
\end{abstract}

\section{Introduction}

Neural networks~\cite{Haykin:1998} provide a functional representation with significant expressive power~\cite{Chen:2014,Schmidhuber:2015,Jordan:2015,Sze:2017}.  Despite concerns about overfitting and lack of reproducibility~\cite{Hutson:2018}, the well-documented successes of neural networks (NNs) are a significant spur for exploration of their utility in the natural sciences, where unique challenges are present~\cite{Hartana:1993,Stewart:2017,Bouzerdoum:1993,Xia:2005,Rudd:2015,Ling:2016,Hennigh:2017,Baker:2019}, including in particular utilization for potential energy surface (PES) representation in chemistry~\cite{Ramakrishnan:2015,Wei:2016,Khorshidi:2016,Liu:2017,Liu:2018,Ferguson:2018,Amabilino:2019,Unke:2021}. There is, of course, a long tradition of PESs using functional representations relying on specific structure involving choices of basis functions~\cite{Sun:1998,Kirschner:2008,Huang:2013,Maier:2015,Chen:2017,Wood:2017,Unke:2021}. Despite the many achievements of these methods, recent years have seen a parade of successes of NNs and machine learning (ML) relative to this established base, demonstrating superior performance and accuracy~\cite{Behler:2008a,Handley:2010,Natarajan:2015,Kolb:2016,Ho:2016,Hellstrom:2017,Behler:2017a,Botu:2017,Yao:2017,Pietrucci:2017,Schutt:2017,Schutt:2017a,Kun:2018,Lubbers:2018,Unke:2021,Bin:2020}. NNPESs offer multiple advantages over classical force fields. The functions used to describe chemical interactions within classical models can limit the types of molecules that the model can describe well, and are rarely able to model chemical reactions that involve breaking or formation of covalent bonds. NNPESs offer much more flexible forms and are limited primarily by the amount and quality of the data used to train them. Further, compared to quantum mechanical methods, NNPESs can offer speed ups of several orders of magnitude.

A key ingredient of a successful NNPES representation is its satisfaction of various necessary symmetries/invariances that are exhibited in molecular structure. Moreover, a highly useful/desirable property is \emph{extensibility}. In other words, a PES representation built from some set of structures ought to be useful for accurate description of the PES of other structures not in the training set, but involving the same set of atoms, charge, and multiplicity. One key element of a PES functional construction that governs its utility is the set of variables used as a feature vector to represent molecular structure. A ground state PES is in principle a function of internal coordinates of all the atoms in the molecule and its charge, and to be sure, many NN based studies rely directly on internal coordinates~\cite{Blank:1995,Gassner:1998,Lorenz:2004,Ho:2016,Rodriguez:2018,Schneider:2017}. Other work also used internal coordinates with Gaussian processes (GPs), rather than NNs~\cite{Stecher:2014}. However, such representations lack extensibility, as the PES representation for a different size molecule would have a different dimension for its input space. Moreover, the dimensionality of this input specification grows very fast with molecule size, and is not optimal, as internal constraints on the coordinates imply the presence of lower dimensional structure.

Lower dimensional representations of molecular structure are often referred to as collective variables (CVs). The choice of symmetry-preserving, fixed-size, extensible CVs, in which a low dimensional PES representation can be built in a wide class of molecules, is challenging~\cite{Chen:2015}. There are available means to discover such good coordinates for any given molecule using purely data-centric methods~\cite{Zheng:2013,Chiavazzo:2017,Georgiou:2017}. However, the enforcement of symmetries, fixed feature size, and extensibility is yet lacking in these methods. There are also recent constructions using local sub-networks, preserving symmetries~\cite{Schutt:2017,Han:2018}, however the extensibility of these methods beyond a limited database of molecules has yet to be demonstrated.

One alternative to the exclusive use of data to construct CVs is to use analytical CV formulations constructed based on chemical intuition and experience, satisfying sought-after properties of fixed-size and symmetry preservation, and to calibrate them based on data. 
There have been many formulations along these lines, e.g. Coulomb matrix~\cite{Rupp:2012,Hansen:2013}, ``bag-of-bonds"~\cite{Hansen:2015}, SMILES strings~\cite{Weininger:1989,Sanchez-Lengeling:2018}, and others~\cite{Bartok:2010,Rogers:2010,Bartok:2013,Duvenaud:2015,Wei:2016,Pham:2018,Hirn:2017,Sidky:2018,Tang:2018}. In particular, the symmetry functions (SFs) of Behler and Parrinello~\cite{Behler:2007}, have found continued development~\cite{Behler:2011,Behler:2015,Behler:2017,Jose:2012}, and extensive successful demonstrations in NN representations of both potential and free energy surfaces~\cite{Cubuk:2017,Khaliullin:2011,Behler:2008,Artrith:2012,Artrith:2014,Galvelis:2017,Onat:2018,Natarajan:2015,Kolb:2016,Smith:2017,Smith:2018}. 
Notably, they have been successfully demonstrated in gas phase systems, showing good accuracy in the representation of quantum mechanical reaction probabilities for multiple reactions~\cite{Kolb:2016}. Behler \& Parrinello SFs, with extensions including atomic-number differentiated atomic environment vectors (AEVs), have been found to provide extensibility in a NN setting for large databases of organic molecules~\cite{Smith:2017}. In this work we employ these AEVs as atomic representations for our model. 

For a potential energy surface to be useful for applications such as optimization and dynamics it needs to be able to predict not only accurate single point energies but also atomic forces. This can be achieved by training solely on single point energies and using enough training data that the model has good accuracy on force predictions. An alternate approach involves training on atomic forces and integrating to get the energies~\cite{Chmiela:2017}. It has in fact been demonstrated that, for an equal number of training points sampled from near minima structures, the inclusion of atomic forces during the training process can yield higher accuracy in both force and energy predictions as compared to energy-only training~\cite{Christensen:2020}. The inclusion of force data improves model predictions away from the training data by ensuring good representation of the energy gradient at the training data points. Christensen and von Lilienfeld~\cite{Christensen:2020} showed that, when training to a variety of molecular sizes and composition, the effects of force training on model performance for potential energy prediction are diminished. However, these models still show a significant improvement on the prediction of atomic forces when compared to their non-force trained counterparts~\cite{Christensen:2020}. 

In the present study, we demonstrate the improvements brought about by force training not only near minima structures but also along a reaction pathway between two isomers. Accurate forces near transition states are of vital importance for applications involving reactions as they are necessary to find the correct transition state as well as the pathway connecting it to the correct minima. Training on atomic forces has seen use in both materials~\cite{Behler:2015,Bartok:2015,Huang:2017,Cooper:2020} and chemistry applications~\cite{Pukrittayakamee:2009,Nguyen:2012,Nguyen:2015,Nandi:2019a,Christensen:2020,Conte:2020,Bowman:2022,Zaverkin:2023,Singraber:2019,Gastegger:2017, Nandi:2019b, Houston:2020, Chen:2020}. Studies include training not just on forces and energies but also on other properties such as dipoles~\cite{Unke:2021b}. Generally, such multimodal training can be advantageous when the various modalities/properties are intrinsically linked (such as forces to energies through the first derivative), but can hurt the accuracy of the model on specific quantities of interest (QoIs) if it is forced to balance the importance of unrelated QoIs. For QoIs that depend on the Hessian matrix, such as infrared spectra, force training can be useful as QoIs are again linked through the derivative~\cite{Gastegger:2017,Herr:2018,Kun:2018}. In this work we focus only on energies and forces of molecules. 

Training on both energies and forces can result in greater memory use per data point/structure, as well as increased training costs due the need for the computation of the second derivative of the molecular energies during the backpropagation process, but, depending on NN construction details, it need not affect the predictive speed of the trained model. Further, the acquisition of forces from many DFT calculations is not much more computationally costly than the acquisition of energies. As might be expected, training to atomic forces seems to be most impactful when a smaller number of training points are available~\cite{Nandi:2019a}. 

We explore the utility of joint energy+force training on the NNPES representation of a portion of the PES of C$_5$H$_5$ spanning two wells connected by a saddle point. The energy difference between the lowest of these wells and the transition state ($\sim$61\,kcal/mol above the lower more stable well) is beyond the scope of most classical molecular dynamic simulations and cannot be modeled properly by most classical force fields due to the breaking and formation of a covalent bond, making it an excellent showcase of the utility of NNPESs. 

We present the relevant details of the construction, including the AEV formulation, the NN construction, and the associated gradients/Jacobians necessary for force-training. Using a computational database of C$_5$H$_5$ structures, which we generate using random mode sampling in the vicinity of the three stationary points and along the intrinsic reaction coordinate connecting them, we train the NNPES and examine its accuracy. In particular, we examine the performance of energy+force versus energy-only training for different training data sizes, and explore the effects of different loss-function force-to-energy weight ratios during training to identify an optimal balance between the two. This ratio was found after normalizing the contribution of both the energy and forces in the loss function to prevent the model from favoring one over the other due simply to the difference in magnitude of the two modalities being learned.

We find that 
the inclusion of force training significantly reduces the test error of the model on both energies and forces. For the largest data set, with 2048 training points, we find an order of magnitude difference between the L2 error of the energy only and energy+force trained models for both forces and energies. Comparing the error of the energy+force trained and energy trained models on molecules generated from normal mode sampling around the three stationary points, we find that the force trained model shows lower error on the majority of structures. We show how this increased accuracy on forces and energy translates to increased accuracy in frequency and zero point energy calculations, and that the force trained model predicts zero point energies within chemical accuracy of the level of theory it was trained to.

In the following, we begin by outlining the problem formulation, including the AEV design, energy and force prediction, as well as NN and loss function construction. We then present highlights of the software implementation and the training data generation. Finally, we proceed to presenting results and discussing their implications.

\section{Problem Formulation}\label{sec:formu}
We begin first with the definition of the feature vector, followed by the NN construction.

\subsection{Design of the Feature Vector}
Consider a configuration of $N$ atoms $\{A_1,\ldots,A_N\}$. Let the Cartesian coordinates of $A_i$ be $x_i\in\RR^3$, and let $\bm{x}=(x_1,\ldots,x_N)\in\RR^{3N}$ be the vector of coordinates of all $N$ atoms. Further, let the set of $n$ atom types in the system be $T=\{T_1,\ldots,T_n\}$, where the type of $A_i$ is defined as the corresponding chemical element $\mathcal{T}_i:=\mathcal{T}(A_i)$. For convenience, we define the configuration as the $N$-tuple $\mathcal{C}=(\mathcal{T}_1,\ldots,\mathcal{T}_N)$. Thus, \emph{e.g.}, the configuration $\mathcal{C}=(\H,\H,\O)$, involves the set of atom types $T=\{\O,\H\}$. Next, let the index set of atoms of type $\tau$ in $\mathcal{C}$ be $S_\tau$, where 
\be
S_\tau = \{i\ |\ \mathcal{T}_i=\tau,\ i=1,\ldots,N\},\quad \tau=T_1,\ldots,T_n .
\ee
Similarly, we define the index set of pair-wise atom indices, as
\bea
S_{\tau,\kappa} &=& \{(j,k)\ |\ \mathcal{T}_j=\tau,\ \mathcal{T}_k=\kappa,\ j=1,\ldots,N-1,\ k=j+1,\ldots,N\},\\
&\qquad& \tau=T_1,\ldots,T_n;\ \kappa=\tau,\ldots,T_n. \nonumber 
\eea  
For example, for a system with $T=\{\C,\H,\O\}$, and the two configurations $(\H,\H,\O)$ and $(\C,\O,\O)$, we have the index sets shown in Table~\ref{tab:configs}. Note that, with $n$ atom types, there are $n$ single-atom index sets $S_\tau$, and $m=n(n+1)/2$ pair-wise index sets $S_{\tau,\kappa}$.

\begin{table}[t]
\small
\centering
\begin{tabular}{|c||c|c|c||c|c|c|c|c|c|}
\hline
$\mathcal{C}$& $S_\C$ & $S_{\H}$ & $S_\O$  & $S_{\C,\C}$ & $S_{\C,\H}$ & $S_{\C,\O}$ & $S_{\H,\H}$ & $S_{\H,\O}$ & $S_{\O,\O}$ \\
\hline
(H,H,O)        & \{\} & \{1,2\} & \{3\} & \{\} & \{\} & \{\} & \{(1,2)\} & \{(1,3),(2,3)\} & \{\}\\
\hline
(C,O,O)        & \{1\} & \{\} & \{2,3\} & \{\} & \{\} & \{(1,2),(1,3)\} & \{\}& \{\} & \{(2,3)\} \\
\hline
\end{tabular}
\caption{Sample configurations and associated index sets with $T=\{\C,\H,\O\}$.}\label{tab:configs}
\end{table}

Following Smith \emph{et al.}~\cite{Smith:2017}, we define the contribution to the potential energy of the system due to $A_i$ as $E_{A_i}(\mathcal{C},\bm{x})$, such that the total potential energy is 
\be
E = \sum_{i=1}^N E_{A_i}(\mathcal{C},\bm{x}).
\label{eq:E}
\ee
We use the ``atomic environment vector" (AEV)~\cite{Smith:2017} as the feature vector that summarizes the geometry of the system in the neighborhood of each atom. We write the AEV of $A_i$ as $y_i(\mathcal{C},\bm{x})\in\RR^M$, and use one NN, $\mathcal{N}_{\mathcal{\tau}}(y):\RR^M\rightarrow\RR$, for each atom type $\tau$, such that 
\be
E = \sum_{i=1}^N \mathcal{N}_{\mathcal{T}_i}(y_i(\mathcal{C},\bm{x})).
\ee
The AEV involves both radial and angular geometry components following Behler and Parinello~\cite{Behler:2007}. For $A_i$, we write the radial components of $y_i$ for pairings with each available atom type separately, according to the $S_\tau$ index sets. Similarly, we write angular components of $y_i$, for groupings with pairs of atom types separately according to the index sets $S_{\tau,\kappa}$. This is done in each case by summing contributions of the pairings/groupings within each set. The construction relies on summarizing geometry information localized around each atom. A key element of the localization involves the cutoff function 
\be
f_c(R_{ij},R_c) = 
\begin{cases}
  0.5 \cos \Big(\frac{\pi R_{ij}}{R_c}\Big) + 0.5 \quad &\mathrm{for}\  R_{ij}\le R_c \\
        0 & \mathrm{otherwise} 
\end{cases}
\ee
where $R_{ij} = \|x_i-x_j\|$ is the Euclidean distance between $(A_i,A_j)$, and $R_c$ is a cutoff radius.
Then with $\mu:=(\eta,\rho)\in\RR^2$, we define the radial components of $y_i$ for pairings with atom type $\tau$ as 
\be
y^{\tau,\mu}_i = \sum_{j\in S_\tau, j\ne i} e^{-\eta(R_{ij}-\rho)^2} \, f_c(R_{ij},R_c^r)
\ee 
with $\mu\in\{\mu_1,\ldots,\mu_{M_r}\}$, and where $R_c^r$ is a radial-SF cutoff radius. 
Further, with $\nu:=(\xi,\gamma,\zeta,\alpha)\in\RR^4$, we define the angular components of $y_i$, for pairings with atom types $(\tau,\kappa)$, as
\be
y^{\tau,\kappa,\nu}_i = \sum_{(j,k)\in S_{\tau,\kappa},\, j\ne i, k\ne i} (0.5+0.5\cos(\theta_{ijk}-\alpha))^\zeta\ 
e^{-\xi(0.5(R_{ij}+R_{ik})-\gamma)^2}\ f_c(R_{ij},R_c^a)\, f_c(R_{ik},R_c^a)
\ee
where $\nu\in\{\nu_1,\ldots,\nu_{M_a}\}$, $R_c^a$ is an angular-SF cutoff radius, and $\theta_{ijk}\in [0,\pi]$ is the angle, centered on $A_i$, between the two vectors $x_{ij}:=x_j-x_i$ and $x_{ik}:=x_k-x_i$, given by
\be
\theta_{ijk} := \arccos \frac{ x_{ij} \cdot x_{ik} } { \|x_{ij}\| \|x_{ik}\|}.
\ee

\begin{figure}
\centerline{
\includegraphics[clip,trim=0mm 0mm 10mm 5mm,angle=0,width=0.5\textwidth]{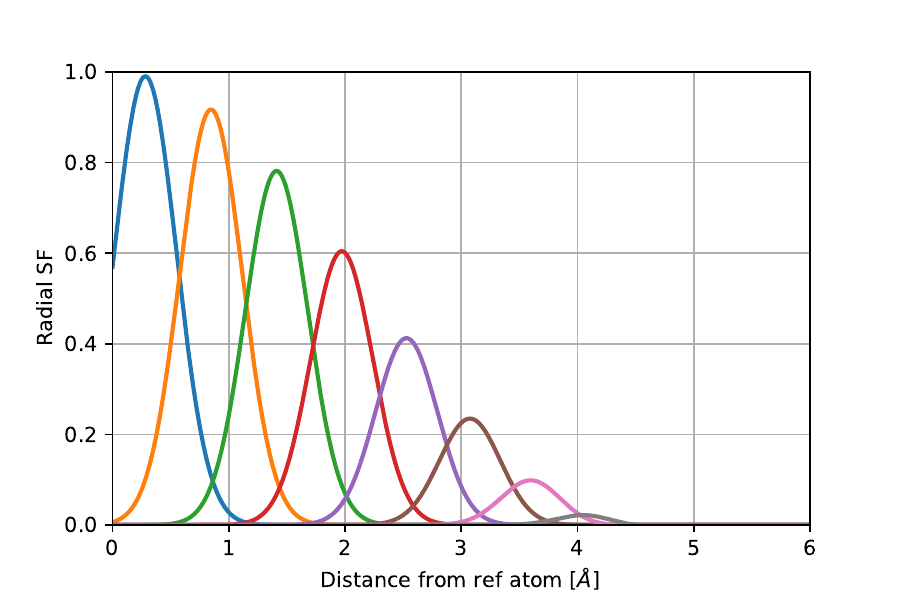}
}
\caption{Radial symmetry functions around a reference atom $A_i$.}
    \label{fig:radSF}
\end{figure}

\begin{figure}
\centerline{
\includegraphics[clip,trim=25mm 0mm 30mm 0mm,angle=0,width=0.3\textwidth]{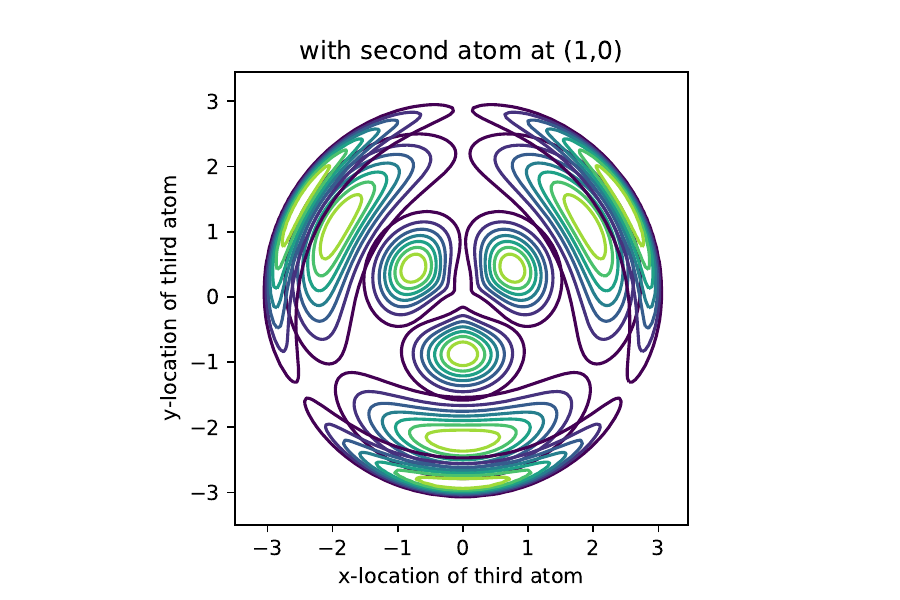}
\includegraphics[clip,trim=25mm 0mm 30mm 0mm,angle=0,width=0.3\textwidth]{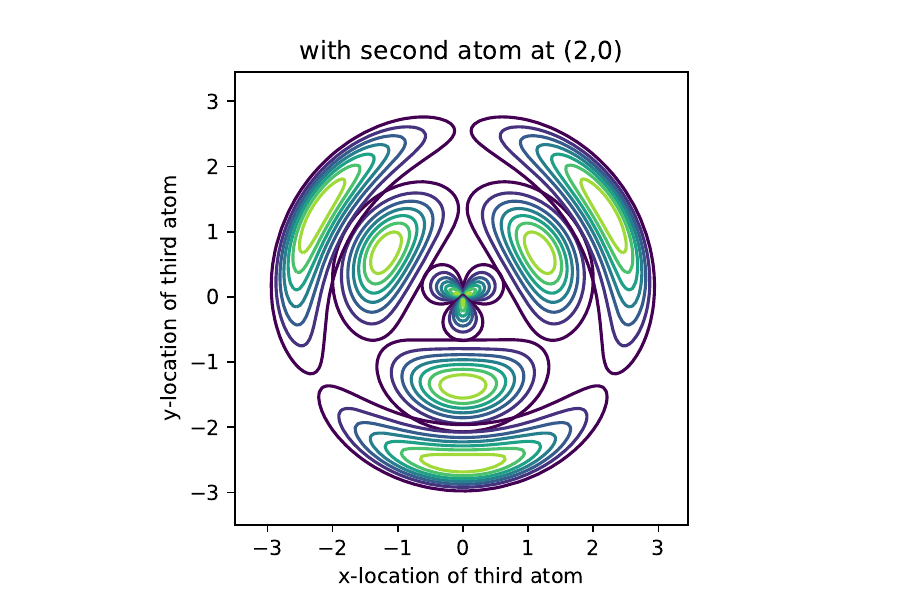}
\includegraphics[clip,trim=25mm 0mm 30mm 0mm,angle=0,width=0.3\textwidth]{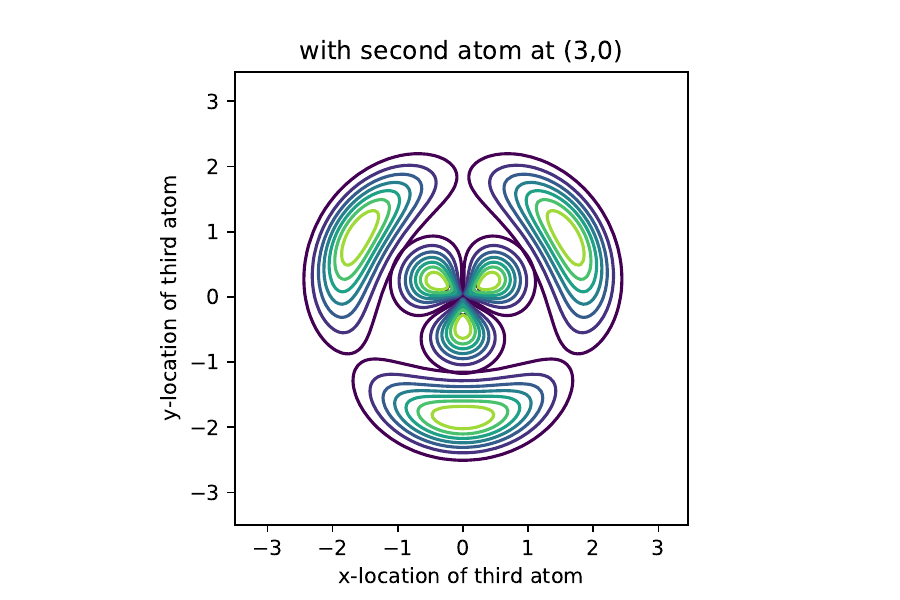}
}
\caption{Angular symmetry functions around a reference atom $A_i$ at $(0,0)$.}
    \label{fig:angSF}
\end{figure}

The full length of the AEV is $M=M_r\, n + M_a\, m$. With 
\bea
y_i^{\tau} &:=& (y_i^{\tau,\mu_1},\ldots,y_i^{\tau,\mu_{M_r}}) \in \RR^{M_r}\\
y_i^{\tau,\kappa} &:=& (y_i^{\tau,\kappa,\nu_1},\ldots,y_i^{\tau,\kappa,\nu_{M_a}}) \in \RR^{M_a}
\eea
we write the full AEV as
\be
y_i = (y^{\tau_1}_i,\ldots,y^{\tau_n}_i,y^{(\tau,\kappa)_1}_i,\ldots,y^{(\tau,\kappa)_m}_i ) 
\in \RR^M.
\ee
Thus, \emph{e.g.}, for $A_i$ in a system with atom types $\{\C,\H\}$, $y_i = ( y^{\C}_{i}, y^{\H}_{i} ,y^{\C,\C}_{i}, y^{\C,\H}_{i}, y^{\H,\H}_{i}) \in \RR^M$, with $M=2M_r+3M_a$. Corresponding representative radial and angular symmetry functions, whose superposition provides the above radial/angular AEV components, are shown in Figs.~\ref{fig:radSF},~\ref{fig:angSF}. Figure~\ref{fig:radSF} illustrates how the radial symmetry functions resolve the local environment around $A_i$ in a number of radial shells, with decaying relevance at larger radii. Similarly Figure~\ref{fig:angSF} shows the resolution of the angular environment around $A_i$ in terms of angles formed at $A_i$ with a second and a third atom. Each frame corresponds to a different location of the second atom, illustrating the multiple radial/angular shells summarizing the localization of the third atom. 

\subsection{AEV Jacobian}
The AEV provides the NN inputs for potential energy representation. This construction is also usable for representation of forces, being the negative spatial gradient of the energy. We present here the extension of the AEV construction for force computations.

We have so far expressed the 3D coordinates of $A_j$ as $x_j$. Here we expressly write the formulation in terms of the 3 spatial coordinates, where we define $x_j:=(x_{j1},x_{j2},x_{j3})$. The PES for a configuration with $N$ atoms is the function $E(x_1,\ldots,x_N)$, being a function on a $3N$-dimensional space, although rotational and translational potential energy invariances imply that the true dependence is generally in a $3N-6$ dimensional space. DFT computations can typically provide both the energy and its gradients with respect to components of $\bm{x}=(x_1,\ldots,x_N)$, namely $\nabla_{x_j} E$, for $j=1,\ldots,N$. We write the gradient operator as
\be
\nabla_{x_j} = \Big(\frac{\partial}{\partial x_{j1}},\frac{\partial}{\partial x_{j2}},\frac{\partial}{\partial x_{j3}}\Big),
\ee
such that,
\be
\nabla_{x_j} E = \sum_{i=1}^N \nabla_{x_j} \mathcal{N}_{\tau_i}(y_i(\mathcal{C},\bm{x})),
\label{eq:gradE}
\ee
where, recalling that $y_i=(y_{i1},\ldots,y_{iM})$, we have
\be
\nabla_{x_j}\mathcal{N}_{\tau_i}(y_i(\mathcal{C},\bm{x})) = \sum_{r=1}^M \frac{\partial\mathcal{N}_{\tau_i}}{\partial y_{ir}} 
\nabla_{x_j} y_{ir} .
\label{eq:gradNN}
\ee
Both the NN and AEV gradients can be computed directly from their analytical constructions. Notably, the AEV Jacobian 
\be
J^{ij}_{rk} = \{\partial y_{ir}/\partial x_{jk}\}
\label{eq:aevJac}
\ee 
where $J^{ij}_{rk}$ is the derivative of the $A_i$ AEV component $r$ with respect to the $A_j$ Cartesian coordinate $k$, is a key element of the construction for force computations. The detailed formulation of the analytical AEV Jacobian is omitted here, but is available elsewhere~\cite{Najm:2021}.

Recall that the AEV, given its particular atom-centered radial/angular construction, is invariant to solid body translation or rotation of the configuration. Further, the additive potential energy construction renders the resultant energy prediction also invariant to permutations of same-type atoms. The enforcement of these physical constraints on the energy representation by construction is an important feature of the AEV-NN representation, ensuring that NN training is constrained in a physical manifold. While the same physical constraints hold for the force vectors in terms of their magnitudes and relative orientations, this is not true for their Cartesian components, specifically as concerns rotational and permutational invariance. This dependence of force Cartesian components on rotation and permutation is indeed captured in the AEV-NNPES representation. The NN-predicted force components are computed as derivatives of the energy function as in Eq.~\ref{eq:gradE}, and thus of the AEV as in Eq.~\ref{eq:gradNN}, with respect to Cartesian coordinates. The NNPES construction captures the rotational and permutational dependencies of the force Cartesian components via the AEV Jacobian. 

\begin{figure}
\centerline{
  \includegraphics[width=0.8\textwidth,clip,trim=0 0 0 0]{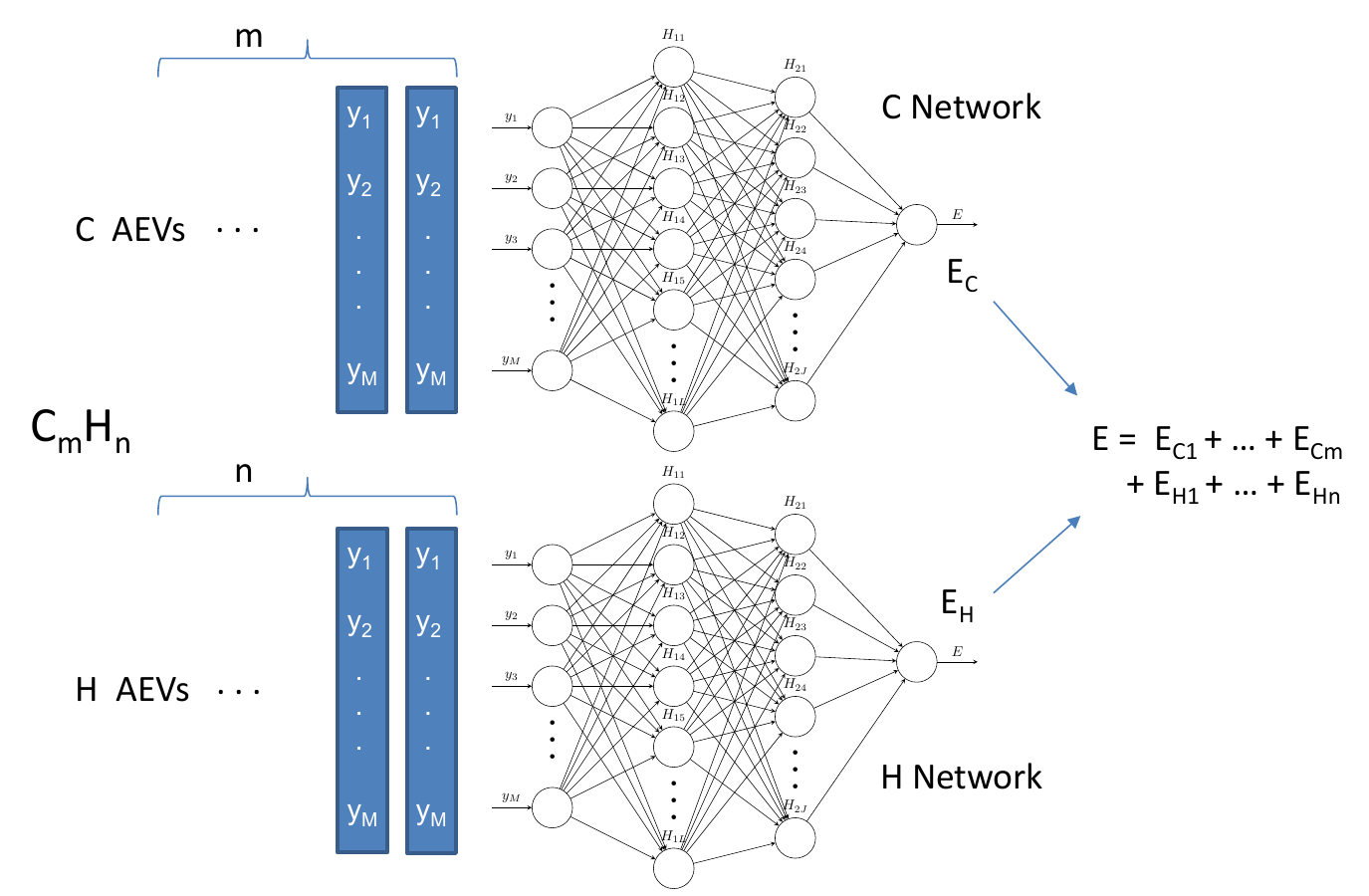}
}
\caption{NNPES schematic for a molecule/configuration involving atom types $(\C,\H)$.}
    \label{fig:nnpes}
\end{figure}

\subsection{Neural Network Construction}\label{sec:nn}
Consider a fully connected feed forward NN with $L$ layers. Let the input be the vector $\bm{u}_0\in\RR^{n_0}$, and the output of layer $\ell=1,\ldots,L$ be $\bm{u}_\ell\in\RR^{n_\ell}$. With the activation function defined as $f_\ell():\RR \rightarrow \RR$, we write the output of each layer as $\bm{u}_\ell=(u_{\ell,1},\ldots,u_{\ell,n_\ell})$, given by
\be
\bm{u}_\ell = \bm{f}_\ell( \bm{W}_\ell \bm{u}_{\ell-1} + \bm{b}_\ell)
\ee
where $\bm{f}_\ell(\bm{q}):=(f_\ell(q_1),\ldots,f_\ell(q_{n_\ell}))$ for $\bm{q}\in\RR^{n_\ell}$, $\bm{W}_\ell
       =\{w_{\ell,ij}\} \in \RR^{n_{\ell}\times n_{\ell-1}}$
and $\bm{b}_\ell\in\RR^{n_\ell}$. 
Then, for any batch of $K$ input vectors, we define the data input matrix $\bm{U}_0=[\bm{u}_{0,1}  \cdots \bm{u}_{0,K}]$
such that
\be
\bm{U}_\ell = \bm{F}_\ell(\bm{W}_\ell\bm{U}_{\ell-1} + \bm{B}_\ell)
\ee
so that $\bm{B}_\ell  = [\bm{b}_{\ell},\ldots,\bm{b}_{\ell}]\in \RR^{n_\ell\times K}$, and, for matrix $\bm{Q}\in \RR^{n_\ell\times K}$, with column vectors $\bm{q}_1, \ldots, \bm{q}_K$,  we have
$
\bm{F}_\ell (\bm{Q}) := [\bm{f}_\ell(\bm{q}_1) \cdots \bm{f}_\ell(\bm{q}_K)].
$
For our purposes, the output node (last layer $L$) is a scalar, thus $n_L=1$, and we employ $f_L(u)\equiv u$.

As indicated above, we use a separate NN for each atom type, employing the same above NN structure for each. With the AEV $y_i$, for $A_i$, provided as input to the NN $\mathcal{N}_{\mathcal{T}_i}$ this construction provides the corresponding NN output as $E_{A_i}$, to be used in Eq.~\ref{eq:E}. Considering our present case with two atom types, Figure~\ref{fig:nnpes} illustrates the overall computation of the potential energy for a molecule composed of $\{\C,\H\}$ atoms using the NNPES construction. Again, the overall detailed formulation is presented elsewhere~\cite{Najm:2021}.

\subsection{Loss Function}\label{sec:loss}
Given a dataset of $K$ configurations/structures where energies $E^D=\{E_1,\ldots,E_K\}$ and forces $\bm{F}^D=\{\bm{F}_1,\ldots,\bm{F}_K\}$ are known from quantum chemistry computations, we quantify the accuracy of the NNPES using a loss function that computes the norm of the error in NNPES predicted quantities $(E,\bm{F})$ compared to their values in the database $(E^D,\bm{F}^D)$. For configuration $k$ with $N_k$ atoms, energy $E_k$, and force vector $\bm{F}_k:=-(\nabla_{x_1}E_k,\ldots,\nabla_{x_{N_k}}E)=(F_{k,1},\ldots,F_{k,3N_k}) \in \RR^{3N_k}$, we define the (mean for forces) square (L2) errors as
\bea
\mathcal{E}^E_k &=& |E_k-E^D_k|^2 \\
\mathcal{E}^F_k &=& \|\bm{F}_k-\bm{F}^D_k\|^2 / (3N_k).
\eea
Further, implementing a weight $\lambda\in[0,1]$, and the normalizing scale factors 
\bea
d_E&=&\max_k(E^D_k)-\min_k(E^D_k)\\
d_F&=&\max_{k,r}(F^D_{k,r})-\min_{k,r}(F^D_{k,r}),
\eea
we define the aggregate mean-square loss as 
\be
\mathcal{E} =  \frac{(1-\lambda)}{K d^2_E} \sum_{k=1}^K \mathcal{E}^E_k + \frac{\lambda}{K d^2_F} \sum_{k=1}^K \mathcal{E}^F_k
\ee
where division by $(d_E,d_F)$ provides normalization of the energy and force magnitudes on $[0,1]$, and the weight $\lambda$ provides relative weighting of energy and force errors. We define the ratio $\lambda/(1-\lambda)$ as the force to energy ratio (F:E) and discuss its effect on training and test errors in the results section.

\section{Software Implementation} \label{sec:software}
We used \textsc{Pytorch}~\cite{pytorch:2019} as a machine learning framework, in a \textsc{Python} context. Given the costs associated with AEV and Jacobian computations, we implement these (using the analytical Jacobian) in \textsc{C++}, relying on \textsc{Pybind11}~\cite{pybind11} for \textsc{Python}-\textsc{C++} binding. We published this AEV code as an open source library \textsc{aevmod}~\cite{aevmod,Najm:2021}. Although not directly relevant here, the use of the resulting NNPES for geometry optimization requires the Hessian of the energy function, and thus of the AEV. We implemented this Hessian similarly in \textsc{C++} in \textsc{aevmod}, relying on automatic differentiation using \textsc{Sacado}~\cite{Sacado}.

\section{Data} \label{sec:data}
\begin{figure}
\centerline{
  \includegraphics[width=0.8\textwidth,clip,trim=0 0 0 0]{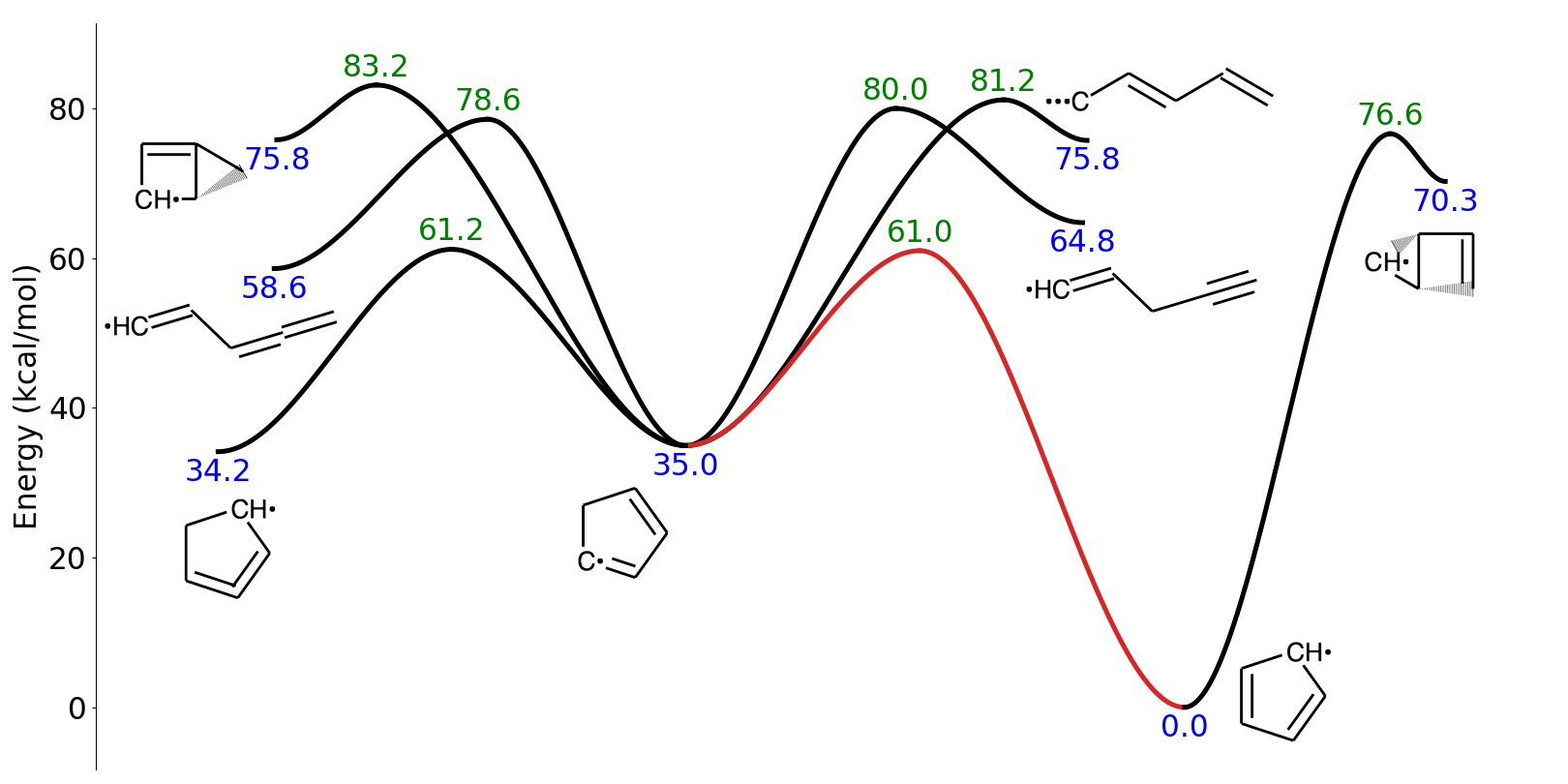}
}
\caption{PES of the reaction being studied (red) and other surrounding reactions (black).}
    \label{fig:c5h5_pes}
\end{figure}

For demonstration, we focus on DFT computations on the \ce{C5H5} PES using QChem~\cite{qchem:web}. Energies and forces are evaluated at the $\omega$B97X-D/6-311++G(d,p) level of theory for a portion of the 24-dimensional potential energy surface. We consider a two-well system, where the two wells are connected via a saddle point. The system is comprised of the lowest energy species on the \ce{C5H5} PES (cyclopentadienyl), the saddle point corresponding to the reaction with the lowest energy barrier cyclopentadienyl can undergo (1,2-H atom shift), and the product resulting from it. The two wells and the associated reaction (highlighted in red) are shown in Fig.~\ref{fig:c5h5_pes}.

We train an ensemble of NNs using subsets of the DFT database.
The training and validation sets for each ensemble member were randomly sampled from the same pool of structures
containing a total of 10875 atomic configurations (25$\times$435). 
The pool of structures used for training, validation, and testing was generated from 25 anchor points, being the three stationary points of the system and 22 points along the intrinsic reaction coordinate (IRC) connecting them. From each of these 25 points, 434 additional structures were generated using randomized normal mode sampling corresponding to $T=2000$\,K. This process involved stretching and compressing the structure at each anchor point along its 24 vibrational modes. We modeled our normal mode sampling based on the sampling method outlined in Smith \emph{et al.}~\cite{Smith:2017}. For each mode $i$, a uniformly distributed pseudo-random number ${\tilde c}_i$ is sampled on the range $[0, 1]$. The ${\tilde c}_i$ are then normalized by their sum and scaled by another uniformly distributed random number $D$ on $[0,1]$, such that
\be\label{eq:Dci}
c_i = \frac{D {\tilde c}_i}{\sum_{j=1}^{3N_a-6} {\tilde c}_j}, \qquad i=1,\ldots,3N_a-6.
\ee
This scaling ensures that $\sum_{i=1}^{3N_a-6} c_i \equiv D \in [0,1]$, which in turn ensures that structures will be randomly sampled at a range of temperatures up to a given maximum corresponding to $D=1$. In this sense, $D$ serves as a measure of the ``normalized temperature distance" from the anchor point to the structure, where, for $D=0$ we have a structure at the anchor point, and for $D=1$ we have a structure at the maximum temperature $T$ above the anchor point. 

However, we made an empirical correction to the normal mode sampling procedure for low-frequency modes. The physical reason for this is that the small force constants of these modes drive the sampling into regions well beyond the validity of a harmonic approximation, yielding extremely high energy points or leading to convergence failures in QChem due to clashing atoms. We designate a frequency $f_i$ low if $f_i<150$~cm$^{-1}$ and replace $\tilde c_i$ with $\tilde c_{i,\mathrm{mod}}$ as
\be
\tilde c_{i,\mathrm{mod}} = \tilde c_i(f_i/150)^4.
\ee
The displacement along mode $i$, $R_i$, is then given by
\be
R_i = \pm\sqrt{\frac{3c_iN_ak_bT}{K_i}},
\ee
where $N_a$ is the number of atoms, $k_b$ is Boltzmann's constant, and $K_i$ is the force constant of the mode. The sign of $R_i$ is chosen randomly from a Bernouli distribution with $p=0.5$ to ensure sampling of both sides of the harmonic potential. The molecule is then perturbed along that mode by its normalized normal mode coordinate scaled by $R_i$. 

\section{Results and discussion}
For each case, we trained an ensemble of neural networks, each with different random initial parameters and training/validation sets. All models were trained with a 256-long AEV with the specific construction parameters outlined in the supplementary information. For each atomic network, we used a fully connected NN architecture of 256:128:64:64:1 with a Gaussian activation function $f_\ell(q)=\exp(-q^2)$ for each of the first three layers and a liner function for the final layer. A test set of 59 structures was used to assess the effects of force training and its accuracy compared to energy only trained models. The test data is held apart, and thus not used in the NN training/validation process. The same test data set is used in all cases. The models trained only to energies are henceforth referred to as the energy trained models and the models trained to both energies and forces as the force trained models, for the sake of brevity. 

\subsection{Neural Network Training}
\begin{figure}
\centerline{
  \includegraphics[scale=0.5,width=0.6\textwidth,clip,trim=0 0 0 0]{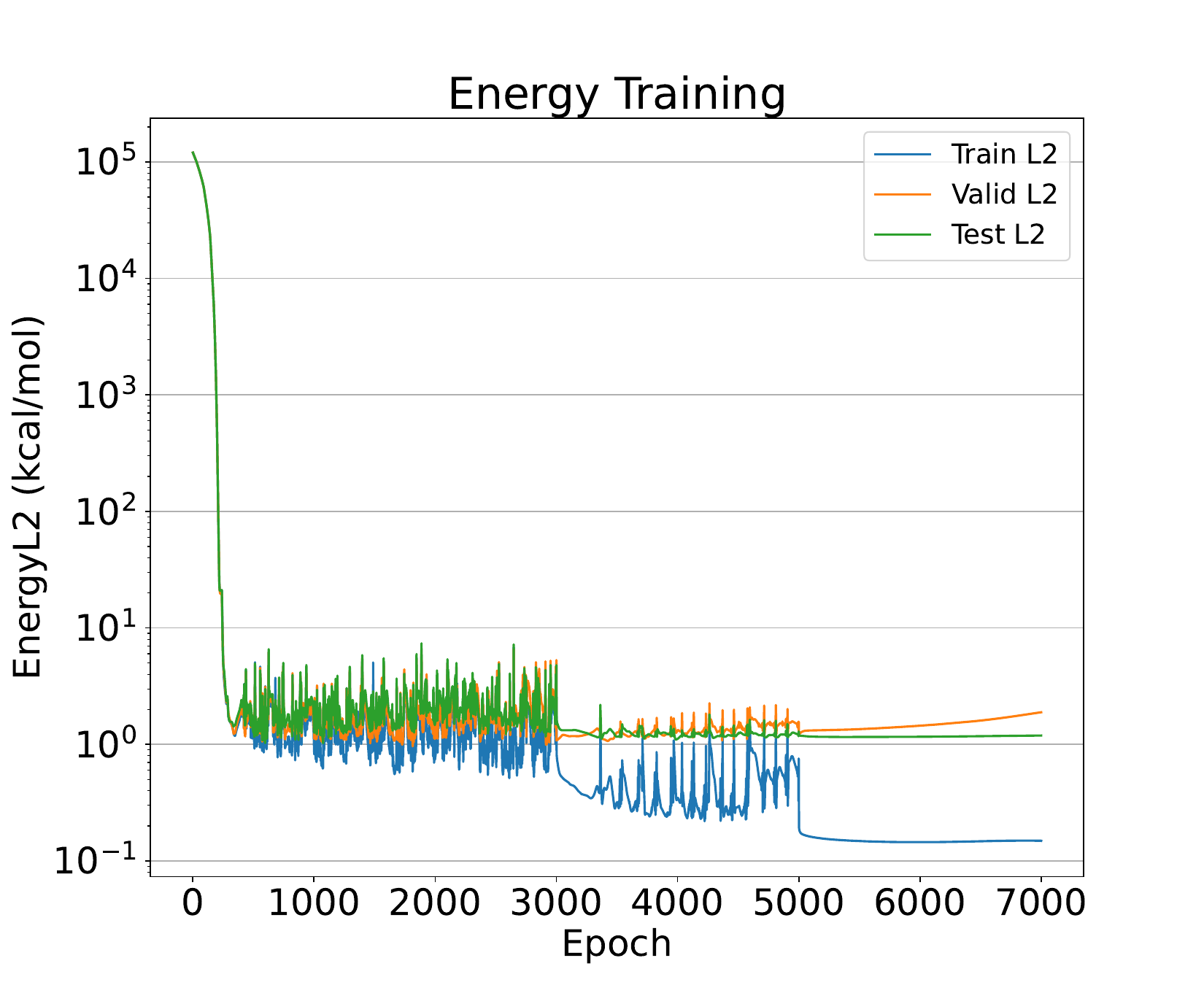}
  \includegraphics[scale=0.5,width=0.6\textwidth,clip,trim=0 0 0 0]{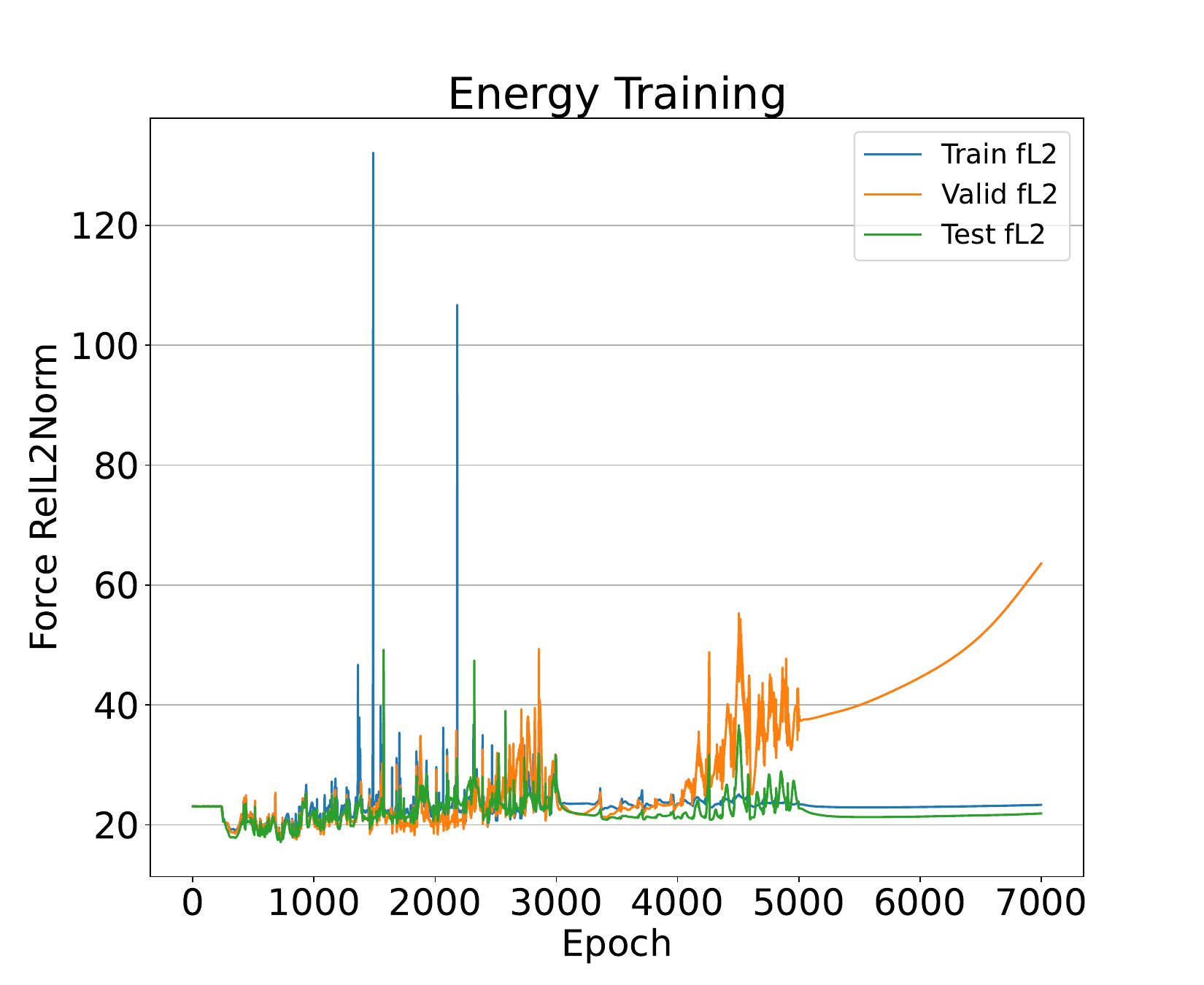}
}
\caption{Plot shows the evolution of the RMSE energy error (left panel) and relative force error (right panel) in the training, testing, and validation set for the energy trained model during training.}
   \label{fig:ETEFep}
\end{figure}

\begin{figure}
\centerline{
  \includegraphics[scale=0.5,width=0.6\textwidth,clip,trim=0 0 0 0]{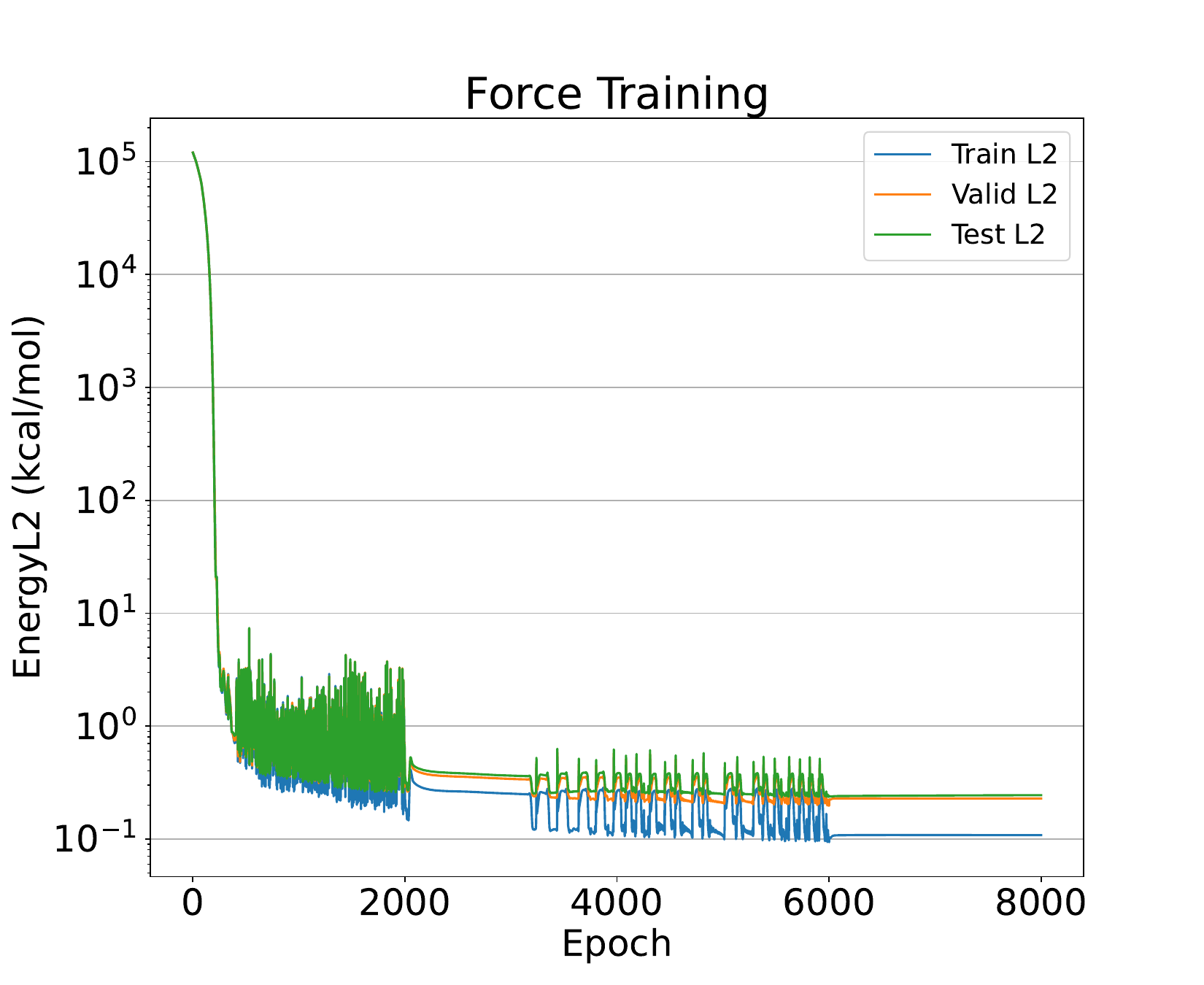}
  \includegraphics[scale=0.5,width=0.6\textwidth,clip,trim=0 0 0 0]{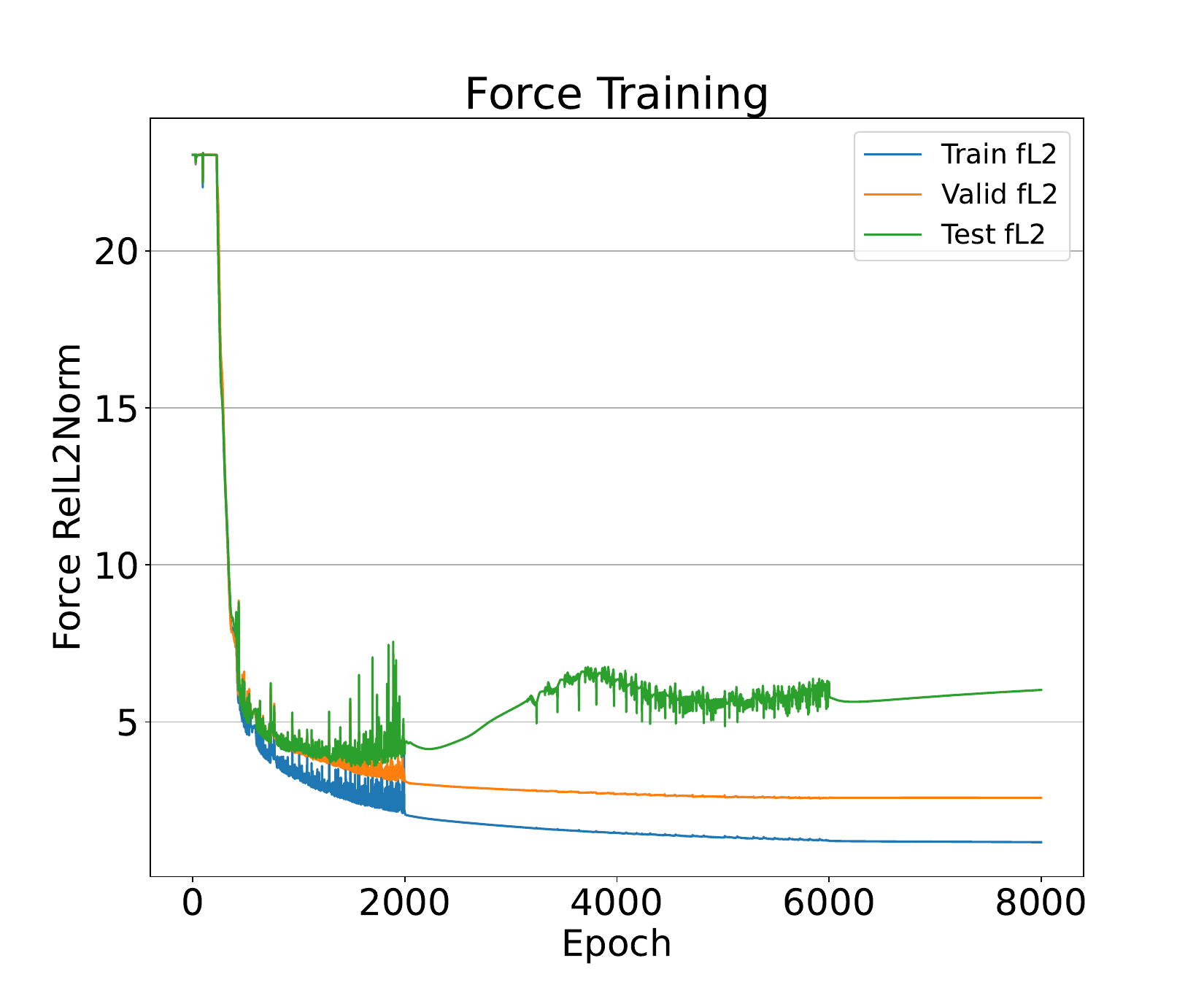}
}
\caption{Plot shows the evolution of the RMSE energy error (left frame) and relative force error (right frame) in the training, testing, and validation set for the force trained model during training.}
   \label{fig:FTEFep}
\end{figure}

For purposes of neural network training, we used a batch size of 8, and made use of a randomized validation set of 64 structures for controlling the learning rate and deciding when to stop training. Each network was trained until its root mean square error (RMSE) on energy for the validation set increased, checking every 1000 epochs. The ADAM algorithm~\cite{kingma:2017} as implemented in \textsc{Pytorch}~\cite{pytorch:2019,pytorch:adam}, was used as the optimizer. The initial learning rate was set to $10^{-3}$. To determine when to update the learning rate, we compared the energy RMSE for the validation and training sets averaged over 100 epochs to those evaluated similarly 1000 epochs earlier. If the change in the average validation RMSE increased while the average training RMSE decreased we multiplied the current learning rate by a factor of 0.1 and continued. Once the learning rate reached our chosen cutoff of $10^{-6}$, this process continued until either the criterion to lower the learning rate was met again or the model had been trained for 50000 epochs, at which point training was stopped. Figures \ref{fig:ETEFep}-\ref{fig:FTEFep} show the evolution of the training/validation/test errors of a single force-trained and energy-trained model with training epoch, each trained to 512 structures. For the force trained model, the energy and forces were given equal weights ($\lambda=0.5$ from equation 22) in the loss function. For each model the energy error rapidly decreased in the early epochs but fluctuated greatly before converging. A similar pattern can be seen for the force error of the force trained model. The force error for the energy trained model begins to decrease after the initial few hundred epochs then fluctuates until converging at a similar level it started at. For both the force and energy errors, the validation error for the model trained only to energy reaches a minimum before increasing in subsequent epochs due to overfitting, given the small size of the data. This does not happen for the force trained models, consistent with the larger amount of information available given the force data. For the subsequent analyses below, unless otherwise stated, the models used to generate the results are those whose parameters were saved at the point of lowest testing error during their training.


\subsection{Weighting Energy and Force Errors in the Loss Function}
\begin{figure}
\centerline{
  \includegraphics[scale=0.2,clip,trim=0 0 0 0]{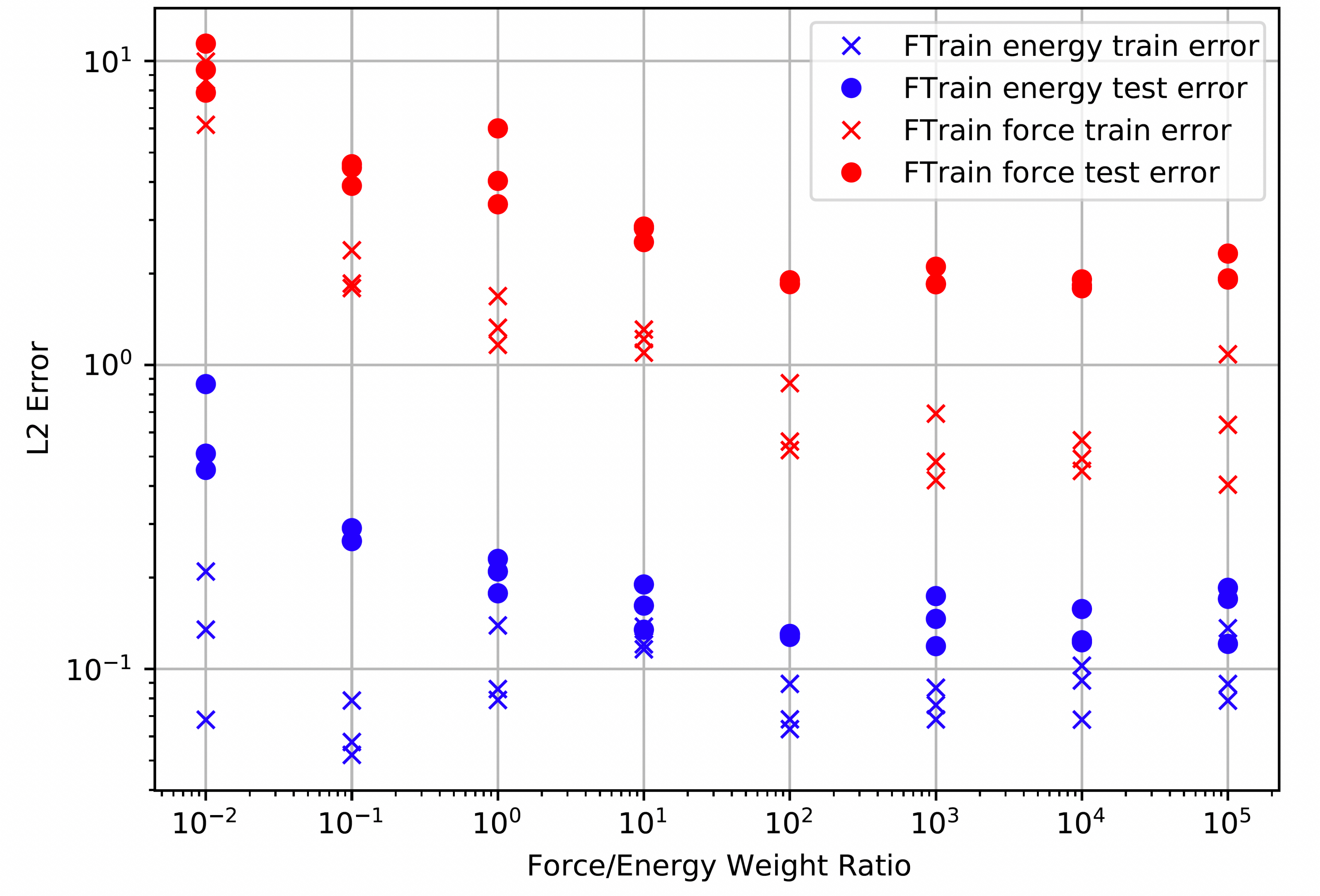}
}
\caption{Training and test errors in energy and forces for energy+force trained models as a function of the F:E ratio. An ensemble of 3 trained models is considered at each F:E ratio condition.}
    \label{fig:xEF_RMSE}
\end{figure}

Finding the optimal ratio of force to energy weight in the loss function is important when training to forces, as the model needs to balance the importance of the two components of the loss function. There are important factors to consider in this regard, such as the difference in scale between the two properties and the amount of data present for each. We discussed earlier above the normalization of the loss for each task in the loss function. This helps to balance the importance of the two quantities even if the absolute scale of the values is different. To properly tune the weighting between the energies and forces for our data set we tested a range of ratios from a $10^{-2}$:1 F:E ratio to a $10^{5}$:1 F:E ratio using 512 structures in the training set. These results are shown in Figure \ref{fig:xEF_RMSE}. We used the root mean square (RMS) norm as a measure of the error of the energy predictions and the relative RMS norm as a measure of the error in force prediction. We define the relative RMS error in the forces as:
\be\label{eq:RRMSEF}
\mathcal{E}_{\mathrm{F}} = \sqrt{\frac{\sum_k \|\bm{F}_k-\bm{F}^D_k\|^2 }{\sum_k \|\bm{F}^D_k\|^2}}
\ee
For each F:E ratio case we trained an ensemble of 3 NNs to provide a sense of the scatter in the results. Both initial weights and training/validation data were randomly sampled for each training ensemble member. We see that going to low F:E ratios does not exhibit a reliable trend in the training error in energy (TrE), with a possible increase in TrE as the F:E ratio is increased to large values, if any trend is to be believed given the scatter. On the other hand, the training error in force (TrF) exhibits a clear upward trend with reduced F:E weight ratio, again without a clear trend in the high F:E limit. Broadly speaking, this is expected, as in the low F:E limit, the model is essentially trained only on energies, thus not providing much of a constraint on the force error. For high F:E weight ratio, the model is trained largely on the forces, with little constraint on the energies, so a possible rise in TrE is conceivable. Clearly, the results indicate that the rise in TrF for low F:E ratio is much more significant than any corresponding trend in TrE for high F:E ratio. Looking at the test error is, we see that the test errors on both energy and force increase significantly in the limit of very low F:E ratio. This is a clear indication of poor generalizability in this limit, highlighting the critical need for force training to ensure good test error, i.e. good accuracy away from the training data. As for test error in the limit of high F:E ratio, the trend, if any is to be inferred is slightly up in both errors. 
Overall, we may say that a ratio of 100:1 F:E in the loss function roughly yielded the lowest error on both the energies and forces on the test set. For the training error no one ratio performed reliably best on both the forces and energies for the range of ratios we used. Interestingly, even for the ratios that most heavily favored forces the error in the energies was always lower than the error in the forces.

Generally, it is fair to say that the optimal weight ratio to use when training for multiple tasks can vary greatly depending on the dataset used and model architecture. A thorough search across many ratios can lead to lower error and therefore a more useful model. Knowledge of the system being studied and the properties of interest can help to narrow the range of values to search but it is still important to perform a search as the factors determining the optimal ratio vary from system to system. A model trained only to near equilibrium points likely has a smaller range of forces and energies in its training set than one trained to points along an IRC and this may affect the ratio that should be used.

\subsection{Dependence of Test Errors on Training Data Size}
\begin{figure}[ht]
\centerline{
  \includegraphics[width=0.6\textwidth,clip,trim=0 0 0 0mm]{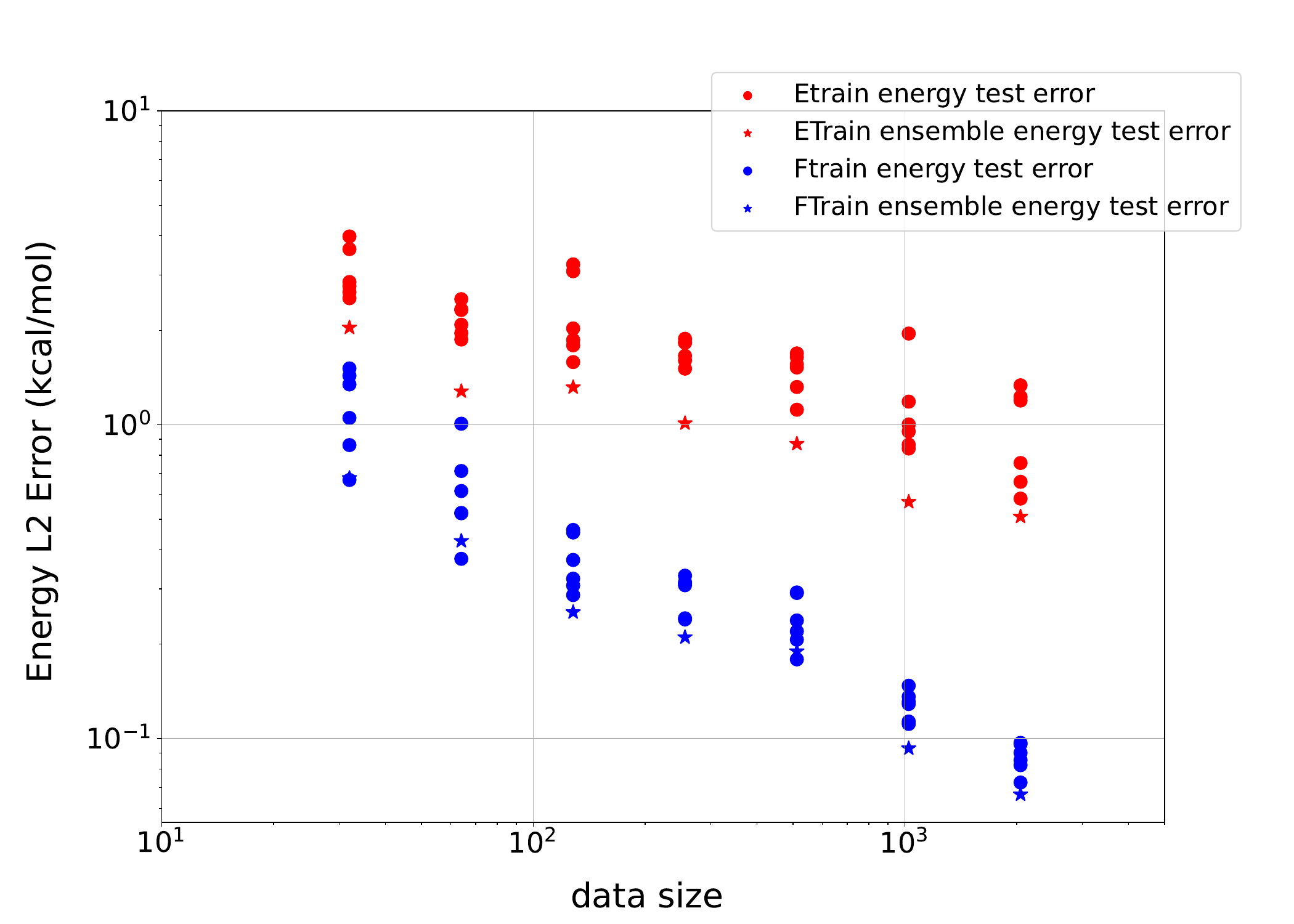}
}
\caption{Plot shows reduction in the RMS error in energy (kcal/mol) with increasing training data size. It also highlights the two cases of energy-only (ETrain) and energy+force (FTrain) training. Results illustrate the lower energy test error in the FTrain case, by about an order of magnitude for 2048 training points.}
\label{fig:E_RMSE}
\end{figure}

\begin{figure}[ht]
\centerline{
  \includegraphics[width=0.6\textwidth,clip,trim=0 0 0 0mm]{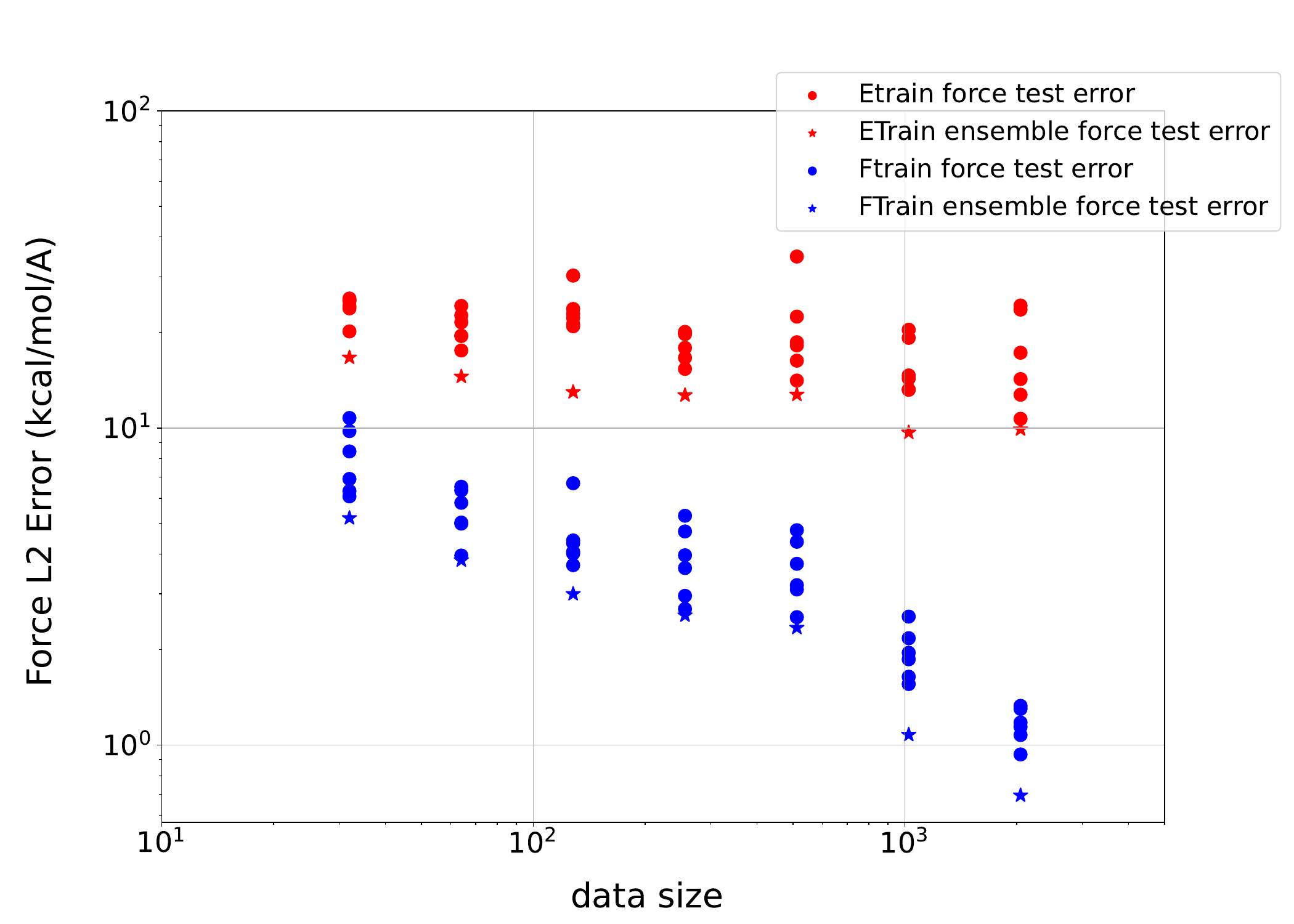}
}
\caption{Plot shows reduction in the RMS error in force (kcal/mol/A) with increasing training data size. It also highlights the two cases of energy-only (ETrain) and energy+force (FTrain) training. Results illustrate the lower force test error in the FTrain case, by about an order of magnitude for 2048 training points.}
\label{fig:F_RMSE}
\end{figure}

Figures (\ref{fig:E_RMSE},\ref{fig:F_RMSE}) show the effect of force training on test error for a range of training set sizes from 32 to 2048 in both energy and force predictions using a F:E ratio of 100:1 in the loss function. We show the test error scatter for an ensemble of 6 NNs, as well as the test error from the ensemble mean model. In all cases every ensemble member of the force trained models outperformed the energy only trained models. Each individual force trained network also outperformed the ensemble average of the energy only trained models for the same training set size for both forces and energies. The test error ratio between the force and energy trained models increased with increasing training set size. This is true for both the forces and energies. For the largest training set, the ratio is about an order of magnitude. The difference in accuracy of the force predictions is even more noticeable. The energy trained  models show little improvement in force prediction with increasing data set size while the error in the force trained models decreases by almost an order of magnitude. As the size of the dataset increases, the amount of data being fitted in the force trained model increases much faster than that for the energy trained model. In the case of these models, trained to \ce{C5H5}, there are 31 times as many labels being used to train the force model compared to the energy trained model (3 forces/atom $\times$ 10 atoms/structure + 1 energy, per structure) so the difference between the amount of training data between the models grows as $30N$ where $N$ is the number of structures in the training set. 

Unfortunately, this extra data, as well as the need for the computation of the second derivative of the energy during the backpropagation step, does lead to increased computational training costs. However, when comparing the error of the energy only trained models to the force trained models, it can be seen that even when training to much larger datasets, the energy-only models struggle to match the accuracy of the force trained models. For the energy error the force trained models trained to only 64 structures show lower error on average than the energy-only models trained to 2048 structures. For the force error, even when training to the largest dataset tested the energy-only model shows higher error than the force trained model trained to only the smallest dataset tested. The fact that training to forces requires a much smaller dataset to achieve the same accuracy as an energy-only trained model can help offset the costs associated with training to forces. It is important, of course, to still ensure that the training set used is diverse enough to cover the region of chemical space being studied, but force training can help the model learn the same region of chemical space with fewer data points. Being able to achieve greater test accuracy with a smaller ensemble, using force training rather than energy only can also help decrease training times as fewer networks are required.

\subsection{Error Dependence on Distance from Anchor Points}
In addition to the 59 structures used in the previous test cases, we also compared energy-only and energy+force trained models on structures generated from normal mode sampling at 2000\,K, using the same method described in the data section above, centered around the three stationary points of the system (well minima and the transition state between them). We used an ensemble of 6 neural nets for each model, each with different random initial parameters and training/validation sets pulled from the same pool of molecules. For this analysis, the models whose parameters were saved at the point of lowest validation error lowest validation error were used. These models were trained to 2048 structures and the force trained model used a 100:1 F:E ratio in its loss function. For each stationary point, 20 structures were generated for each vibrational mode for a total of 480 structures per stationary point. Some of these structures failed to converge during the DFT calculations, such that a total of 1423 points (out of 1440) were used for this test. We compared the errors of the energy/force-trained models when predicting energies and forces of these structures. For the energy we show the absolute difference in the prediction of the models and DFT. For the forces we show the relative RMS error as defined above in Eq.~\ref{eq:RRMSEF}. The parity plots in Figures \ref{fig:Eerr_f_vs_e} and \ref{fig:Ferr_f_vs_e} show the performance of each model on each point for energy and forces respectively. 
They also show the normalized temperature distance ($D$ from Eq.~\ref{eq:Dci}) of each structure away from its anchor point. Points below the black diagonal line are structures where the force trained model shows a lower error than the energy trained model, while points above the diagonal exhibit the reverse relationship.

Both models tend to show higher energy error on points further from the anchor points. However, even for these far points, the force trained model performs better on the majority of structures. The correlation between the model test error and the distance from the anchor point is more obvious for the force trained model than the energy trained model in Figure \ref{fig:Ferr_f_vs_e} but for both models the highest error structures are structures that are far from the anchor points. In Figure \ref{fig:Eerr_f_vs_e} there are 1188 structures below the black line and 235 above it. Additionally, the force trained model shows chemical accuracy on 82\% of the points, (1180 out of 1423). The energy trained model achieves this accuracy on only 46\% of the data (650 out of 1423) In Figure \ref{fig:Ferr_f_vs_e} there are 849 structures below the black line and 574 above it. The force trained model shows a relative error as low as $10^{-2}$ on many structures, including the anchor points, while the energy trained model does not achieve even an error of less than of $10^{-1}$ on any of the tested structures. This means that force trained model can be expected to show errors ~1\% of the magnitude of the total forces on the anchor points while the energy trained model will be an order of magnitude worse. Despite showing lower error on most of the points, the highest errors in force and energy prediction for the force trained model are worse than for the energy-trained model.  However, these outliers represent only a small percentage of the total number of points. For both the force trained and energy trained models their worst points show places where the model failed to learn and the results for each model are qualitatively incorrect. The three anchor points (shown as stars) are also included in both graphs as the darkest blue points with a distance of zero, although they were also included in the training set for each model. The force trained model outperforms the energy only trained model on each of these three points for both energy and forces.

\subsection{Frequency prediction}
\begin{figure}
\centerline{
    \includegraphics[scale=0.4,clip,trim=0 0 0 0]{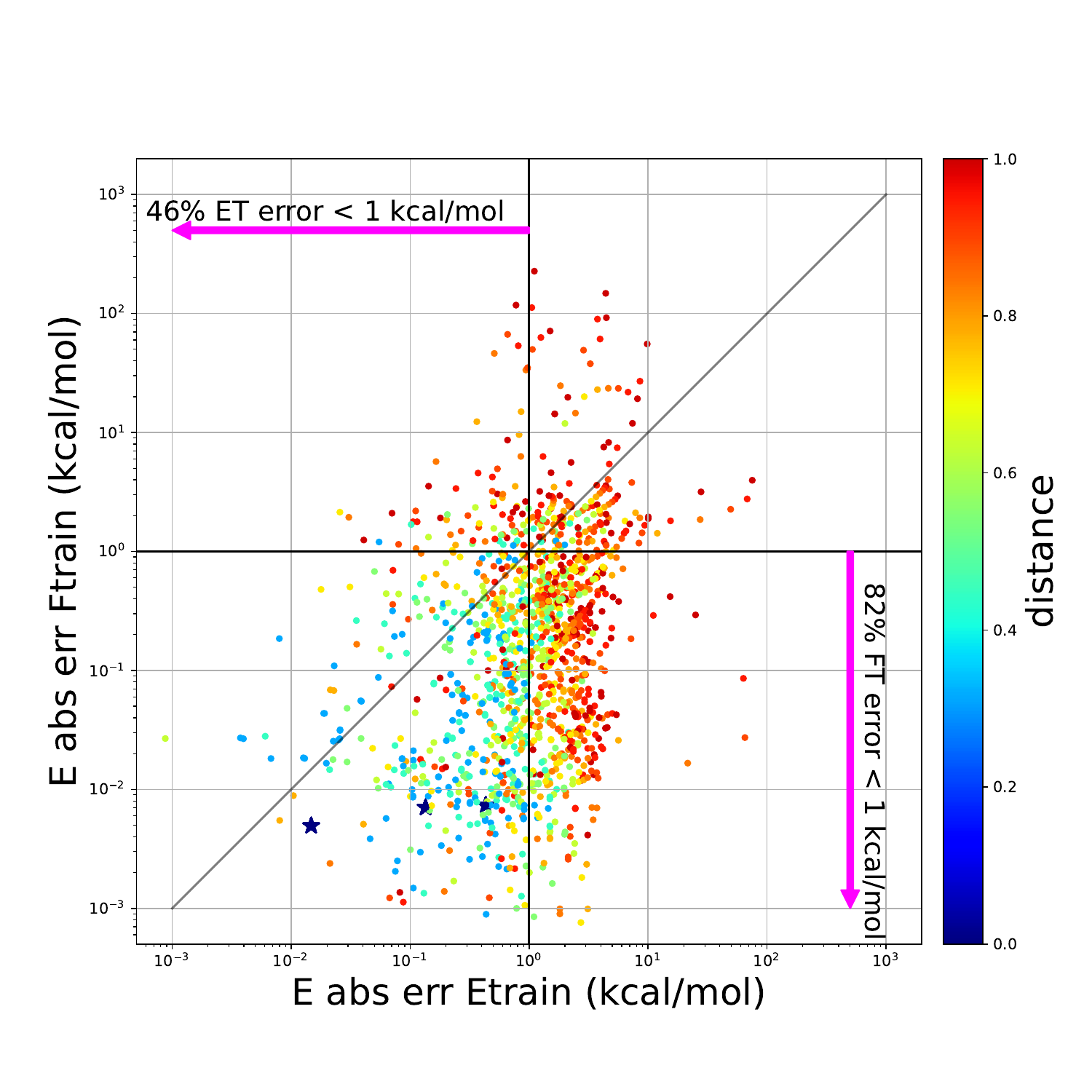}
}
\caption{Energy absolute test error for mean 6-NN predictions at normal mode samples. Plot shows mean prediction test errors resulting from energy+force training versus those from energy training, with 2048 training points. Points are colored according to their normalized distance from anchor points ($D$), where normalization is done separately for each normal mode. The three anchor points at distance 0 are shown as stars. The percentage of points with less than 1.0 kcal/mol error from each model are also shown. 1423 points are plotted in total.} 
    \label{fig:Eerr_f_vs_e}
\end{figure}

\begin{figure}
\centerline{
  \includegraphics[scale=0.4,clip,trim=0 0 0 0]{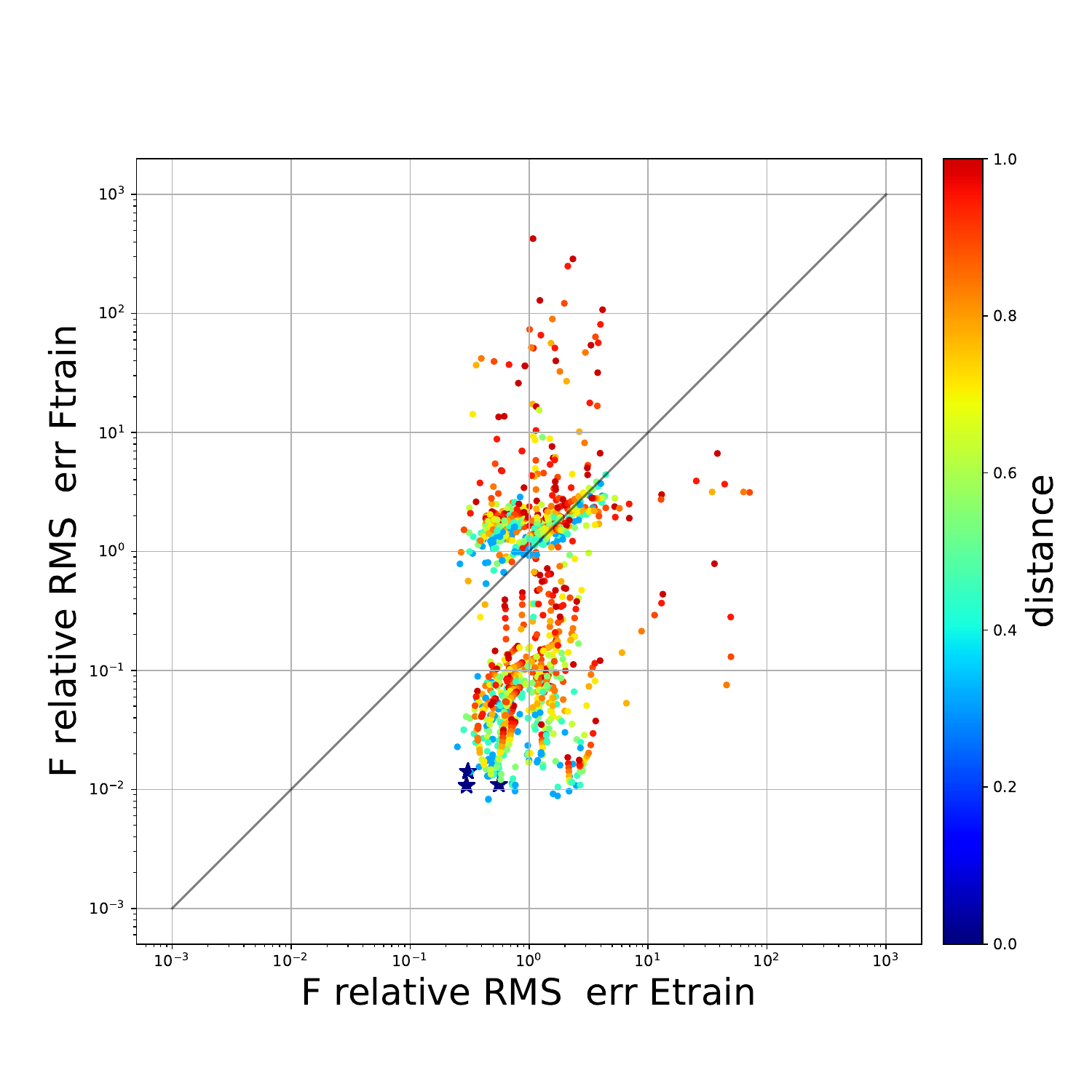}
}
        \caption{Force relative RMS test error for mean 6-NN predictions at normal mode samples. Plot shows mean prediction relative test errors resulting from energy+force training versus those from energy training, with 2048 training points. Points are colored according to their normalized distance from anchor points, where normalization is done separately for each normal mode. The three anchor points at distance 0 are shown as stars.}
    \label{fig:Ferr_f_vs_e}
\end{figure}

\begin{figure}
\centerline{
  \includegraphics[scale=0.3,clip,trim=0 0 0 13.7mm]{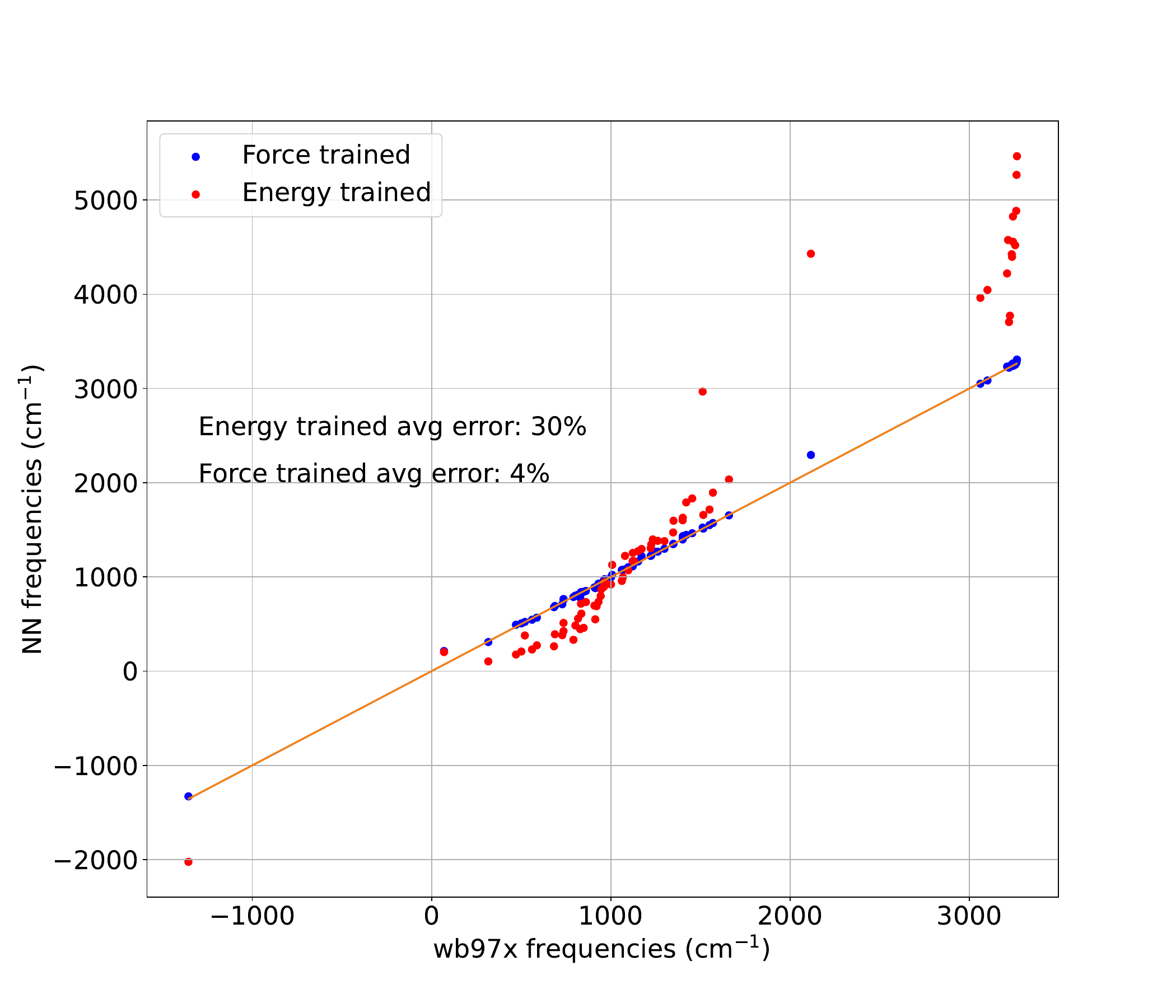}
}
\caption{Frequencies predicted by the force trained(blue) and energy trained(red) models versus frequencies predicted by $\omega$B97X-D/6-311++G(d,p) for each stationary point.}
    \label{fig:FRcor}
\end{figure}

For a final test of the advantages of force training versus energy training and to measure the usefulness of each model on a kinetically relevant property, we computed the vibrational frequencies of each stationary point using both models and compared the results to $\omega$B97X-D/6-311++G(d,p). The same models used to generate figures \ref{fig:Ferr_f_vs_e} and \ref{fig:Eerr_f_vs_e} were used in this comparison. For each model, we optimized the stationary points, starting from the geometry predicted by $\omega$B97X-D/6-311++G(d,p), computed the vibrational frequencies, and the average percent error of the frequencies compared to $\omega$B97X-D/6-311++G(d,p). These results are shown in Figure \ref{fig:FRcor}. The force trained model shows excellent agreement with DFT across the entire range of frequencies and has an average percent error of 4\%. The energy trained model underestimates most of the frequencies below 1000 cm$^{-1}$ and overestimates most of the ones above 1000 cm$^{-1}$, showing an average percent error of $30\%$. Both models correctly optimize the transition state to a transition state, finding only one imaginary frequency, but the energy trained model misses the value of this frequency by several hundred wavenumbers. The zero point energies (ZPE) of the force trained model agree within 0.5 kcal/mol with that of the DFT values, while the ZPE arising from the energy trained model overestimates the ZPE by 6.9, 5.4 and 13.6  kcal/mol for the two wells and the saddle point (respectively), causing a large error in the final, chemically relevant reverse and forward barrier heights, despite the small error in the electronic energy. 

\section{Conclusions} \label{sec:conclusions}

Our results indicate that, with adequate training, a neural network potential energy surface can provide a level of accuracy over reactive landscapes that is comparable to QM methods, but at a much lower predictive computational cost. Of course, the generation of a training set that adequately describes the system of interest can still be a difficult and costly process. Training to forces in addition to energy can greatly decrease the amount of training data needed without sacrificing model accuracy. Training to both energies and forces increased the accuracy of the model for both energy and force prediction. A force trained model trained on a small set of structures shows lower error than a model trained to an order of magnitude more structures without force data. However, the trained model accuracy does depend on the relative weights given to the forces versus the energies in the loss function. The models trained only to energy show slower improvement in force prediction with increasing training set size as compared to force trained models. Force training also results in greater accuracy when predicting kinetically relevant properties than training to energy alone. The force trained model shows excellent correlation with DFT when predicting vibrational frequencies, and its zero point energy predictions are more than an order of magnitude lower in error than the energy trained model.

We believe force training can allow for larger and more complex systems to be studied using NNPESs, systems which would otherwise be too costly to generate enough training data for to achieve requisite accuracy from energy-only training. Further, the cost of running DFT calculations grows as $\mathcal{O}(N^3)$, making the creation of large training sets for large molecules a costly process. Force training allows the same level of accuracy to be achieved with a much smaller training set so fewer of these costly calculations are required.  

\section{Acknowledgement}
This work was supported by the US Department of Energy (DOE), Office of Basic Energy Sciences (BES) Division of Chemical Sciences, Geosciences, and
Biosciences, under contract number 022187. 
This research used resources of the National Energy Research Scientific Computing Center (NERSC), a DOE Office of Science User Facility supported by the Office of Science of the U.S. DOE under Contract No. DE-AC02-05CH11231 using NERSC Award No. BES-ERCAP0021664. This article has been authored by employees of National Technology \& Engineering Solutions of Sandia, LLC under contract no. DE-NA0003525 with the U.S. Department of Energy (DOE). The employees co-own right, title, and interest in and to the article and are responsible for its contents. The United States Government retains and the publisher, by accepting the article for publication, acknowledges that the United States Government retains a nonexclusive, paid-up, irrevocable, worldwide license to publish or reproduce the published form of this article or allow others to do so, for United States Government purposes. The DOE will provide public access to these results of federally sponsored research in accordance with the DOE Public Access Plan https://www.energy.gov/downloads/doe-public-access-plan.

\begin{suppinfo}

We specify the various parameters of the \{C,H\} AEV construction as follows. We use

\begin{itemize}
\item Cutoff radius for radial SF: $R_c^r=4.6$\AA
\item Cutoff radius for angular SF: $R_c^a=3.1$\AA 
\item Discretization of $\eta$ uses $n_\eta=1$, $\eta=1.0/\delta_r^2$, with $\delta_r=2h_r/3$, $h_r=R^r_c/n_{\rho}$
\item Discretization of $\rho$ uses $n_{\rho}=32$, $\rho_i=(i-\frac{1}{2})h_r$, $i=1,\ldots,n_{\rho}$
\item We have $M_r = n_\eta n_\rho = 32$, where $(\eta,\rho)\in \RR^{n_\eta}\times\RR^{n_\rho}$ 
\item Discretization of $\xi$ uses $n_\xi = 1$, $\xi=1.0/\delta_a^2$, $\delta_a=2h_a/3$, $h_a=R^a_c/n_{\gamma}$
\item Discretization of $\gamma$ uses $n_{\gamma}=8$, $\gamma_i=(i-\frac{1}{2})h_a$, $i=1,\ldots,n_{\gamma}$
\item Discretization of $\zeta$ uses $n_\zeta=1$, $\zeta=8$
\item Discretization of $\alpha$ uses $n_\alpha=8$, $h=\pi/(n_\alpha-1)$, $\alpha_i= (i-1)h$, $i=1,\ldots,n_\alpha$
\item We have $M_a = n_\xi n_\gamma n_\zeta n_\alpha = 64$, where $(\xi,\gamma,\zeta,\alpha) \in   \RR^{n_\xi}\times\RR^{n_\gamma} \times \RR^{n_\zeta}\times\RR^{n_\alpha}$ 
\item $M=2M_r+3M_a=256$
\end{itemize}

\end{suppinfo}

\bibliography{ref}

\providecommand{\latin}[1]{#1}
\makeatletter
\providecommand{\doi}
  {\begingroup\let\do\@makeother\dospecials
  \catcode`\{=1 \catcode`\}=2 \doi@aux}
\providecommand{\doi@aux}[1]{\endgroup\texttt{#1}}
\makeatother
\providecommand*\mcitethebibliography{\thebibliography}
\csname @ifundefined\endcsname{endmcitethebibliography}
  {\let\endmcitethebibliography\endthebibliography}{}
\begin{mcitethebibliography}{109}
\providecommand*\natexlab[1]{#1}
\providecommand*\mciteSetBstSublistMode[1]{}
\providecommand*\mciteSetBstMaxWidthForm[2]{}
\providecommand*\mciteBstWouldAddEndPuncttrue
  {\def\EndOfBibitem{\unskip.}}
\providecommand*\mciteBstWouldAddEndPunctfalse
  {\let\EndOfBibitem\relax}
\providecommand*\mciteSetBstMidEndSepPunct[3]{}
\providecommand*\mciteSetBstSublistLabelBeginEnd[3]{}
\providecommand*\EndOfBibitem{}
\mciteSetBstSublistMode{f}
\mciteSetBstMaxWidthForm{subitem}{(\alph{mcitesubitemcount})}
\mciteSetBstSublistLabelBeginEnd
  {\mcitemaxwidthsubitemform\space}
  {\relax}
  {\relax}

\bibitem[Haykin(1998)]{Haykin:1998}
Haykin,~S. \emph{Neural Networks: A Comprehensive Foundation}, 2nd ed.;
  Prentice Hall PTR: Upper Saddle River, NJ, USA, 1998\relax
\mciteBstWouldAddEndPuncttrue
\mciteSetBstMidEndSepPunct{\mcitedefaultmidpunct}
{\mcitedefaultendpunct}{\mcitedefaultseppunct}\relax
\EndOfBibitem
\bibitem[Chen and Lin(2014)Chen, and Lin]{Chen:2014}
Chen,~X.-W.; Lin,~X. Big Data Deep Learning: Challenges and Perspectives.
  \emph{IEEE Access} \textbf{2014}, \emph{2}, 514--525\relax
\mciteBstWouldAddEndPuncttrue
\mciteSetBstMidEndSepPunct{\mcitedefaultmidpunct}
{\mcitedefaultendpunct}{\mcitedefaultseppunct}\relax
\EndOfBibitem
\bibitem[Schmidhuber(2015)]{Schmidhuber:2015}
Schmidhuber,~J. Deep learning in neural networks: An overview. \emph{Neural
  Networks} \textbf{2015}, \emph{61}, 85--117\relax
\mciteBstWouldAddEndPuncttrue
\mciteSetBstMidEndSepPunct{\mcitedefaultmidpunct}
{\mcitedefaultendpunct}{\mcitedefaultseppunct}\relax
\EndOfBibitem
\bibitem[Jordan and Mitchell(2015)Jordan, and Mitchell]{Jordan:2015}
Jordan,~M.~I.; Mitchell,~T.~M. Machine learning: trends, perspectives, and
  prospects. \emph{Science} \textbf{2015}, \emph{349}, 255--260\relax
\mciteBstWouldAddEndPuncttrue
\mciteSetBstMidEndSepPunct{\mcitedefaultmidpunct}
{\mcitedefaultendpunct}{\mcitedefaultseppunct}\relax
\EndOfBibitem
\bibitem[Sze \latin{et~al.}(2017)Sze, Chen, Yang, and Emer]{Sze:2017}
Sze,~V.; Chen,~Y.-H.; Yang,~T.-J.; Emer,~J.~S. Efficient Processing of Deep
  Neural Networks: A Tutorial and Survey. \emph{Proceedings of the IEEE}
  \textbf{2017}, \emph{105}, 2295--2329\relax
\mciteBstWouldAddEndPuncttrue
\mciteSetBstMidEndSepPunct{\mcitedefaultmidpunct}
{\mcitedefaultendpunct}{\mcitedefaultseppunct}\relax
\EndOfBibitem
\bibitem[Hutson(2018)]{Hutson:2018}
Hutson,~M. Artificial intelligence faces reproducibility crisis. \emph{Science}
  \textbf{2018}, \emph{359}, 725--726\relax
\mciteBstWouldAddEndPuncttrue
\mciteSetBstMidEndSepPunct{\mcitedefaultmidpunct}
{\mcitedefaultendpunct}{\mcitedefaultseppunct}\relax
\EndOfBibitem
\bibitem[Hartana and Richards(1993)Hartana, and Richards]{Hartana:1993}
Hartana,~R.~K.; Richards,~G.~G. Constrained neural network-based identification
  of harmonic sources. \emph{IEEE Transactions on Industry Applications}
  \textbf{1993}, \emph{29}, 202--208\relax
\mciteBstWouldAddEndPuncttrue
\mciteSetBstMidEndSepPunct{\mcitedefaultmidpunct}
{\mcitedefaultendpunct}{\mcitedefaultseppunct}\relax
\EndOfBibitem
\bibitem[Stewart and Ermon(2017)Stewart, and Ermon]{Stewart:2017}
Stewart,~R.; Ermon,~S. Label-Free Supervision of Neural Networks with Physics
  and Domain Knowledge. AAAI. 2017; pp 2576--2582\relax
\mciteBstWouldAddEndPuncttrue
\mciteSetBstMidEndSepPunct{\mcitedefaultmidpunct}
{\mcitedefaultendpunct}{\mcitedefaultseppunct}\relax
\EndOfBibitem
\bibitem[Bouzerdoum and Pattison(1993)Bouzerdoum, and
  Pattison]{Bouzerdoum:1993}
Bouzerdoum,~A.; Pattison,~T.~R. Neural network for quadratic optimization with
  bound constraints. \emph{IEEE transactions on neural networks} \textbf{1993},
  \emph{4}, 293--304\relax
\mciteBstWouldAddEndPuncttrue
\mciteSetBstMidEndSepPunct{\mcitedefaultmidpunct}
{\mcitedefaultendpunct}{\mcitedefaultseppunct}\relax
\EndOfBibitem
\bibitem[Xia \latin{et~al.}(2005)Xia, Feng, and Wang]{Xia:2005}
Xia,~Y.~S.; Feng,~G.; Wang,~J. A primal-dual neural network for online
  resolving constrained kinematic redundancy in robot motion control.
  \emph{IEEE Transactions on Systems, Man, and Cybernetics, Part B
  (Cybernetics)} \textbf{2005}, \emph{35}, 54--64\relax
\mciteBstWouldAddEndPuncttrue
\mciteSetBstMidEndSepPunct{\mcitedefaultmidpunct}
{\mcitedefaultendpunct}{\mcitedefaultseppunct}\relax
\EndOfBibitem
\bibitem[Rudd and Ferrari(2015)Rudd, and Ferrari]{Rudd:2015}
Rudd,~K.; Ferrari,~S. A constrained integration (CINT) approach to solving
  partial differential equations using artificial neural networks.
  \emph{Neurocomputing} \textbf{2015}, \emph{155}, 277 -- 285\relax
\mciteBstWouldAddEndPuncttrue
\mciteSetBstMidEndSepPunct{\mcitedefaultmidpunct}
{\mcitedefaultendpunct}{\mcitedefaultseppunct}\relax
\EndOfBibitem
\bibitem[Ling \latin{et~al.}(2016)Ling, Jones, and Templeton]{Ling:2016}
Ling,~J.; Jones,~R.; Templeton,~J. Machine learning strategies for systems with
  invariance properties. \emph{Journal of Computational Physics} \textbf{2016},
  \emph{318}, 22--35\relax
\mciteBstWouldAddEndPuncttrue
\mciteSetBstMidEndSepPunct{\mcitedefaultmidpunct}
{\mcitedefaultendpunct}{\mcitedefaultseppunct}\relax
\EndOfBibitem
\bibitem[{Hennigh}(2017)]{Hennigh:2017}
{Hennigh},~O. {Lat-Net: Compressing Lattice Boltzmann Flow Simulations using
  Deep Neural Networks}. \emph{ArXiv e-prints} \textbf{2017}, \relax
\mciteBstWouldAddEndPunctfalse
\mciteSetBstMidEndSepPunct{\mcitedefaultmidpunct}
{}{\mcitedefaultseppunct}\relax
\EndOfBibitem
\bibitem[Baker \latin{et~al.}(2019)Baker, Alexander, Bremer, Hagberg,
  Kevrekidis, Najm, Parashar, Patra, Sethian, Wild, and Willcox]{Baker:2019}
Baker,~N.; Alexander,~F.; Bremer,~T.; Hagberg,~A.; Kevrekidis,~Y.; Najm,~H.;
  Parashar,~M.; Patra,~A.; Sethian,~J.; Wild,~S.; Willcox,~K. Workshop Report
  on Basic Research Needs for Scientific Machine Learning: Core Technologies
  for Artificial Intelligence. \textbf{2019}, {\blue
  https://www.osti.gov/servlets/purl/1478744}\relax
\mciteBstWouldAddEndPuncttrue
\mciteSetBstMidEndSepPunct{\mcitedefaultmidpunct}
{\mcitedefaultendpunct}{\mcitedefaultseppunct}\relax
\EndOfBibitem
\bibitem[Ramakrishnan \latin{et~al.}(2015)Ramakrishnan, Dral, Rupp, and von
  Lilienfeld]{Ramakrishnan:2015}
Ramakrishnan,~R.; Dral,~P.~O.; Rupp,~M.; von Lilienfeld,~O.~A. Big Data Meets
  Quantum Chemistry Approximations: The $\Delta$-Machine Learning Approach.
  \emph{Journal of Chemical Theory and Computation} \textbf{2015}, \emph{11},
  2087--2096, PMID: 26574412\relax
\mciteBstWouldAddEndPuncttrue
\mciteSetBstMidEndSepPunct{\mcitedefaultmidpunct}
{\mcitedefaultendpunct}{\mcitedefaultseppunct}\relax
\EndOfBibitem
\bibitem[Wei \latin{et~al.}(2016)Wei, Duvenaud, and Aspuru-Guzik]{Wei:2016}
Wei,~J.~N.; Duvenaud,~D.; Aspuru-Guzik,~A. Neural Networks for the Prediction
  of Organic Chemistry Reactions. \emph{ACS Central Science} \textbf{2016},
  \emph{2}, 725--732\relax
\mciteBstWouldAddEndPuncttrue
\mciteSetBstMidEndSepPunct{\mcitedefaultmidpunct}
{\mcitedefaultendpunct}{\mcitedefaultseppunct}\relax
\EndOfBibitem
\bibitem[Khorshidi and Peterson(2016)Khorshidi, and Peterson]{Khorshidi:2016}
Khorshidi,~A.; Peterson,~A.~A. {Amp: A modular approach to machine learning in
  atomistic simulations}. \emph{Computer Physics Communications} \textbf{2016},
  \emph{207}, 310--324\relax
\mciteBstWouldAddEndPuncttrue
\mciteSetBstMidEndSepPunct{\mcitedefaultmidpunct}
{\mcitedefaultendpunct}{\mcitedefaultseppunct}\relax
\EndOfBibitem
\bibitem[Liu \latin{et~al.}(2017)Liu, Ramsundar, Kawthekar, Shi, Gomes,
  Luu~Nguyen, Ho, Sloane, Wender, and Pande]{Liu:2017}
Liu,~B.; Ramsundar,~B.; Kawthekar,~P.; Shi,~J.; Gomes,~J.; Luu~Nguyen,~Q.;
  Ho,~S.; Sloane,~J.; Wender,~P.; Pande,~V. Retrosynthetic Reaction Prediction
  Using Neural Sequence-to-Sequence Models. \emph{ACS Central Science}
  \textbf{2017}, \emph{3}, 1103--1113\relax
\mciteBstWouldAddEndPuncttrue
\mciteSetBstMidEndSepPunct{\mcitedefaultmidpunct}
{\mcitedefaultendpunct}{\mcitedefaultseppunct}\relax
\EndOfBibitem
\bibitem[Liu \latin{et~al.}(2018)Liu, Zhou, Zhou, Zhang, Luo, Guo, and
  Jiang]{Liu:2018}
Liu,~Q.; Zhou,~X.; Zhou,~L.; Zhang,~Y.; Luo,~X.; Guo,~H.; Jiang,~B.
  Constructing High-Dimensional Neural Network Potential Energy Surfaces for
  Gas–Surface Scattering and Reactions. \emph{The Journal of Physical
  Chemistry C} \textbf{2018}, \emph{122}, 1761--1769\relax
\mciteBstWouldAddEndPuncttrue
\mciteSetBstMidEndSepPunct{\mcitedefaultmidpunct}
{\mcitedefaultendpunct}{\mcitedefaultseppunct}\relax
\EndOfBibitem
\bibitem[Ferguson(2018)]{Ferguson:2018}
Ferguson,~A.~L. ACS Central Science Virtual Issue on Machine Learning.
  \emph{ACS Central Science} \textbf{2018}, \emph{4}, 938--941\relax
\mciteBstWouldAddEndPuncttrue
\mciteSetBstMidEndSepPunct{\mcitedefaultmidpunct}
{\mcitedefaultendpunct}{\mcitedefaultseppunct}\relax
\EndOfBibitem
\bibitem[Amabilino \latin{et~al.}(2019)Amabilino, Bratholm, Bennie, Vaucher,
  Reiher, and Glowacki]{Amabilino:2019}
Amabilino,~S.; Bratholm,~L.~A.; Bennie,~S.~J.; Vaucher,~A.~C.; Reiher,~M.;
  Glowacki,~D.~R. Training Neural Nets to Learn Reactive Potential Energy
  Surfaces using Interactive Quantum Chemistry in Virtual Reality. \emph{The
  Journal of Physical Chemistry A} \textbf{2019}, \emph{0}, null, PMID:
  30892040\relax
\mciteBstWouldAddEndPuncttrue
\mciteSetBstMidEndSepPunct{\mcitedefaultmidpunct}
{\mcitedefaultendpunct}{\mcitedefaultseppunct}\relax
\EndOfBibitem
\bibitem[Unke \latin{et~al.}(2021)Unke, Chmiela, Sauceda, Gastegger, Poltavsky,
  Schütt, Tkatchenko, and Müller]{Unke:2021}
Unke,~O.~T.; Chmiela,~S.; Sauceda,~H.~E.; Gastegger,~M.; Poltavsky,~I.;
  Schütt,~K.~T.; Tkatchenko,~A.; Müller,~K.-R. Machine Learning Force Fields.
  \emph{Chemical Reviews} \textbf{2021}, \emph{121}, 10142--10186, PMID:
  33705118\relax
\mciteBstWouldAddEndPuncttrue
\mciteSetBstMidEndSepPunct{\mcitedefaultmidpunct}
{\mcitedefaultendpunct}{\mcitedefaultseppunct}\relax
\EndOfBibitem
\bibitem[Sun(1998)]{Sun:1998}
Sun,~H. {}. \emph{J. Phys. Chem. B} \textbf{1998}, \emph{102}, 7338\relax
\mciteBstWouldAddEndPuncttrue
\mciteSetBstMidEndSepPunct{\mcitedefaultmidpunct}
{\mcitedefaultendpunct}{\mcitedefaultseppunct}\relax
\EndOfBibitem
\bibitem[Kirschner \latin{et~al.}(2008)Kirschner, Yongye, Tschampel, no,
  Daniels, Foley, and Woods]{Kirschner:2008}
Kirschner,~K.; Yongye,~A.; Tschampel,~S.; no,~J. G.-O.; Daniels,~C.; Foley,~B.;
  Woods,~R. {}. \emph{J. Comput. Chem.} \textbf{2008}, \emph{29}, 622\relax
\mciteBstWouldAddEndPuncttrue
\mciteSetBstMidEndSepPunct{\mcitedefaultmidpunct}
{\mcitedefaultendpunct}{\mcitedefaultseppunct}\relax
\EndOfBibitem
\bibitem[Huang and MacKerell~Jr(2013)Huang, and MacKerell~Jr]{Huang:2013}
Huang,~J.; MacKerell~Jr,~A.~D. CHARMM36 all-atom additive protein force field:
  Validation based on comparison to NMR data. \emph{Journal of Computational
  Chemistry} \textbf{2013}, \emph{34}, 2135--2145\relax
\mciteBstWouldAddEndPuncttrue
\mciteSetBstMidEndSepPunct{\mcitedefaultmidpunct}
{\mcitedefaultendpunct}{\mcitedefaultseppunct}\relax
\EndOfBibitem
\bibitem[Maier \latin{et~al.}(2015)Maier, Martinez, Kasavajhala, Wickstrom,
  Hauser, and Simmerling]{Maier:2015}
Maier,~J.; Martinez,~C.; Kasavajhala,~K.; Wickstrom,~L.; Hauser,~K.;
  Simmerling,~C. {}. \emph{J. Chem. Theory Comput.} \textbf{2015}, \emph{11},
  3696\relax
\mciteBstWouldAddEndPuncttrue
\mciteSetBstMidEndSepPunct{\mcitedefaultmidpunct}
{\mcitedefaultendpunct}{\mcitedefaultseppunct}\relax
\EndOfBibitem
\bibitem[Chen \latin{et~al.}(2017)Chen, Deng, Tran, Tang, Chu, and
  Ong]{Chen:2017}
Chen,~C.; Deng,~Z.; Tran,~R.; Tang,~H.; Chu,~I.-H.; Ong,~S. {Accurate Force
  Field for Molybdenum by Machine Learning Large Materials Data}. \emph{Phys.
  Rev. Materials} \textbf{2017}, \emph{1}, 043603\relax
\mciteBstWouldAddEndPuncttrue
\mciteSetBstMidEndSepPunct{\mcitedefaultmidpunct}
{\mcitedefaultendpunct}{\mcitedefaultseppunct}\relax
\EndOfBibitem
\bibitem[Wood and Thompson(2017)Wood, and Thompson]{Wood:2017}
Wood,~M.~A.; Thompson,~A.~P. https://arxiv.org/abs/1702.07042\relax
\mciteBstWouldAddEndPuncttrue
\mciteSetBstMidEndSepPunct{\mcitedefaultmidpunct}
{\mcitedefaultendpunct}{\mcitedefaultseppunct}\relax
\EndOfBibitem
\bibitem[Behler \latin{et~al.}(2008)Behler, Marto\v{n}\'ak, Donadio, and
  Parrinello]{Behler:2008a}
Behler,~J.; Marto\v{n}\'ak,~R.; Donadio,~D.; Parrinello,~M. Pressure-induced
  phase transitions in silicon studied by neural network-based metadynamics
  simulations. \emph{physica status solidi (b)} \textbf{2008}, \emph{245},
  2618--2629\relax
\mciteBstWouldAddEndPuncttrue
\mciteSetBstMidEndSepPunct{\mcitedefaultmidpunct}
{\mcitedefaultendpunct}{\mcitedefaultseppunct}\relax
\EndOfBibitem
\bibitem[Handley and Popelier(2010)Handley, and Popelier]{Handley:2010}
Handley,~C.~M.; Popelier,~P. L.~A. Potential Energy Surfaces Fitted by
  Artificial Neural Networks. \emph{The Journal of Physical Chemistry A}
  \textbf{2010}, \emph{114}, 3371--3383\relax
\mciteBstWouldAddEndPuncttrue
\mciteSetBstMidEndSepPunct{\mcitedefaultmidpunct}
{\mcitedefaultendpunct}{\mcitedefaultseppunct}\relax
\EndOfBibitem
\bibitem[Kondati~Natarajan \latin{et~al.}(2015)Kondati~Natarajan, Morawietz,
  and Behler]{Natarajan:2015}
Kondati~Natarajan,~S.; Morawietz,~T.; Behler,~J. Representing the
  potential-energy surface of protonated water clusters by high-dimensional
  neural network potentials. \emph{Phys. Chem. Chem. Phys.} \textbf{2015},
  \emph{17}, 8356--8371\relax
\mciteBstWouldAddEndPuncttrue
\mciteSetBstMidEndSepPunct{\mcitedefaultmidpunct}
{\mcitedefaultendpunct}{\mcitedefaultseppunct}\relax
\EndOfBibitem
\bibitem[Kolb \latin{et~al.}(2016)Kolb, Zhao, Li, Jiang, and Guo]{Kolb:2016}
Kolb,~B.; Zhao,~B.; Li,~J.; Jiang,~B.; Guo,~H. Permutation invariant potential
  energy surfaces for polyatomic reactions using atomistic neural networks.
  \emph{The Journal of Chemical Physics} \textbf{2016}, \emph{144},
  224103\relax
\mciteBstWouldAddEndPuncttrue
\mciteSetBstMidEndSepPunct{\mcitedefaultmidpunct}
{\mcitedefaultendpunct}{\mcitedefaultseppunct}\relax
\EndOfBibitem
\bibitem[Ho \latin{et~al.}(2016)Ho, Pham-Tran, Kawazoe, and Le]{Ho:2016}
Ho,~T.~H.; Pham-Tran,~N.-N.; Kawazoe,~Y.; Le,~H.~M. Ab Initio Investigation of
  O–H Dissociation from the Al–OH2 Complex Using Molecular Dynamics and
  Neural Network Fitting. \emph{The Journal of Physical Chemistry A}
  \textbf{2016}, \emph{120}, 346--355, PMID: 26741404\relax
\mciteBstWouldAddEndPuncttrue
\mciteSetBstMidEndSepPunct{\mcitedefaultmidpunct}
{\mcitedefaultendpunct}{\mcitedefaultseppunct}\relax
\EndOfBibitem
\bibitem[Hellstr{\"o}m and Behler(2017)Hellstr{\"o}m, and
  Behler]{Hellstrom:2017}
Hellstr{\"o}m,~M.; Behler,~J. Structure of aqueous NaOH solutions: insights
  from neural-network-based molecular dynamics simulations. \emph{Phys. Chem.
  Chem. Phys.} \textbf{2017}, \emph{19}, 82--96\relax
\mciteBstWouldAddEndPuncttrue
\mciteSetBstMidEndSepPunct{\mcitedefaultmidpunct}
{\mcitedefaultendpunct}{\mcitedefaultseppunct}\relax
\EndOfBibitem
\bibitem[Behler(2017)]{Behler:2017a}
Behler,~J. First Principles Neural Network Potentials for Reactive Simulations
  of Large Molecular and Condensed Systems. \emph{Angewandte Chemie
  International Edition} \textbf{2017}, \emph{56}, 12828--12840\relax
\mciteBstWouldAddEndPuncttrue
\mciteSetBstMidEndSepPunct{\mcitedefaultmidpunct}
{\mcitedefaultendpunct}{\mcitedefaultseppunct}\relax
\EndOfBibitem
\bibitem[Botu \latin{et~al.}(2017)Botu, Batra, Chapman, and
  Ramprasad]{Botu:2017}
Botu,~V.; Batra,~R.; Chapman,~J.; Ramprasad,~R. Machine Learning Force Fields:
  Construction, Validation, and Outlook. \emph{The Journal of Physical
  Chemistry C} \textbf{2017}, \emph{121}, 511--522\relax
\mciteBstWouldAddEndPuncttrue
\mciteSetBstMidEndSepPunct{\mcitedefaultmidpunct}
{\mcitedefaultendpunct}{\mcitedefaultseppunct}\relax
\EndOfBibitem
\bibitem[Yao \latin{et~al.}(2017)Yao, Herr, Brown, and Parkhill]{Yao:2017}
Yao,~K.; Herr,~J.~E.; Brown,~S.~N.; Parkhill,~J. Intrinsic Bond Energies from a
  Bonds-in-Molecules Neural Network. \emph{The Journal of Physical Chemistry
  Letters} \textbf{2017}, \emph{8}, 2689--2694, PMID: 28573865\relax
\mciteBstWouldAddEndPuncttrue
\mciteSetBstMidEndSepPunct{\mcitedefaultmidpunct}
{\mcitedefaultendpunct}{\mcitedefaultseppunct}\relax
\EndOfBibitem
\bibitem[Pietrucci(2017)]{Pietrucci:2017}
Pietrucci,~F. Strategies for the exploration of free energy landscapes: Unity
  in diversity and challenges ahead. \emph{Reviews in Physics} \textbf{2017},
  \emph{2}, 32 -- 45\relax
\mciteBstWouldAddEndPuncttrue
\mciteSetBstMidEndSepPunct{\mcitedefaultmidpunct}
{\mcitedefaultendpunct}{\mcitedefaultseppunct}\relax
\EndOfBibitem
\bibitem[Sch{\"{u}}tt \latin{et~al.}(2017)Sch{\"{u}}tt, Arbabzadah, Chmiela,
  M{\"{u}}ller, and Tkatchenko]{Schutt:2017}
Sch{\"{u}}tt,~K.~T.; Arbabzadah,~F.; Chmiela,~S.; M{\"{u}}ller,~K.~R.;
  Tkatchenko,~A. {Quantum-chemical insights from deep tensor neural networks}.
  \emph{Nature Communications} \textbf{2017}, \emph{8}, 13890\relax
\mciteBstWouldAddEndPuncttrue
\mciteSetBstMidEndSepPunct{\mcitedefaultmidpunct}
{\mcitedefaultendpunct}{\mcitedefaultseppunct}\relax
\EndOfBibitem
\bibitem[Sch\"{u}tt \latin{et~al.}(2017)Sch\"{u}tt, Kindermans, Sauceda,
  Chmiela, Tkatchenko, and M\"{u}ller]{Schutt:2017a}
Sch\"{u}tt,~K.~T.; Kindermans,~P.-J.; Sauceda,~H.~E.; Chmiela,~S.;
  Tkatchenko,~A.; M\"{u}ller,~K.-R. SchNet: A Continuous-filter Convolutional
  Neural Network for Modeling Quantum Interactions. Proceedings of the 31st
  International Conference on Neural Information Processing Systems. USA, 2017;
  pp 992--1002\relax
\mciteBstWouldAddEndPuncttrue
\mciteSetBstMidEndSepPunct{\mcitedefaultmidpunct}
{\mcitedefaultendpunct}{\mcitedefaultseppunct}\relax
\EndOfBibitem
\bibitem[Yao \latin{et~al.}(2018)Yao, Herr, Toth, Mckintyre, and
  Parkhill]{Kun:2018}
Yao,~K.; Herr,~J.~E.; Toth,~D.~W.; Mckintyre,~R.; Parkhill,~J. The
  TensorMol-0.1 model chemistry: a neural network augmented with long-range
  physics. \emph{Chem. Sci.} \textbf{2018}, \emph{9}, 2261--2269\relax
\mciteBstWouldAddEndPuncttrue
\mciteSetBstMidEndSepPunct{\mcitedefaultmidpunct}
{\mcitedefaultendpunct}{\mcitedefaultseppunct}\relax
\EndOfBibitem
\bibitem[Lubbers \latin{et~al.}(2018)Lubbers, Smith, and Barros]{Lubbers:2018}
Lubbers,~N.; Smith,~J.~S.; Barros,~K. Hierarchical modeling of molecular
  energies using a deep neural network. \emph{The Journal of Chemical Physics}
  \textbf{2018}, \emph{148}, 241715\relax
\mciteBstWouldAddEndPuncttrue
\mciteSetBstMidEndSepPunct{\mcitedefaultmidpunct}
{\mcitedefaultendpunct}{\mcitedefaultseppunct}\relax
\EndOfBibitem
\bibitem[Jiang \latin{et~al.}(2020)Jiang, Li, and Guo]{Bin:2020}
Jiang,~B.; Li,~J.; Guo,~H. High-Fidelity Potential Energy Surfaces for
  Gas-Phase and Gas–Surface Scattering Processes from Machine Learning.
  \emph{The Journal of Physical Chemistry Letters} \textbf{2020}, \emph{11},
  5120--5131\relax
\mciteBstWouldAddEndPuncttrue
\mciteSetBstMidEndSepPunct{\mcitedefaultmidpunct}
{\mcitedefaultendpunct}{\mcitedefaultseppunct}\relax
\EndOfBibitem
\bibitem[Blank \latin{et~al.}(1995)Blank, Brown, Calhoun, and
  Doren]{Blank:1995}
Blank,~T.; Brown,~S.; Calhoun,~A.; Doren,~D. {Neural network models of
  potential energy surfaces}. \emph{J. Chem. Phys.} \textbf{1995}, \emph{103},
  4129\relax
\mciteBstWouldAddEndPuncttrue
\mciteSetBstMidEndSepPunct{\mcitedefaultmidpunct}
{\mcitedefaultendpunct}{\mcitedefaultseppunct}\relax
\EndOfBibitem
\bibitem[Gassner \latin{et~al.}(1998)Gassner, Probst, Lauenstein, and
  Hermansson]{Gassner:1998}
Gassner,~H.; Probst,~M.; Lauenstein,~A.; Hermansson,~K. Representation of
  Intermolecular Potential Functions by Neural Networks. \emph{The Journal of
  Physical Chemistry A} \textbf{1998}, \emph{102}, 4596--4605\relax
\mciteBstWouldAddEndPuncttrue
\mciteSetBstMidEndSepPunct{\mcitedefaultmidpunct}
{\mcitedefaultendpunct}{\mcitedefaultseppunct}\relax
\EndOfBibitem
\bibitem[Lorenz \latin{et~al.}(2004)Lorenz, Groß, and Scheffler]{Lorenz:2004}
Lorenz,~S.; Groß,~A.; Scheffler,~M. Representing high-dimensional
  potential-energy surfaces for reactions at surfaces by neural networks.
  \emph{Chemical Physics Letters} \textbf{2004}, \emph{395}, 210 -- 215\relax
\mciteBstWouldAddEndPuncttrue
\mciteSetBstMidEndSepPunct{\mcitedefaultmidpunct}
{\mcitedefaultendpunct}{\mcitedefaultseppunct}\relax
\EndOfBibitem
\bibitem[Rodriguez \latin{et~al.}(2018)Rodriguez, d’Errico, Facco, and
  Laio]{Rodriguez:2018}
Rodriguez,~A.; d’Errico,~M.; Facco,~E.; Laio,~A. Computing the Free Energy
  without Collective Variables. \emph{Journal of Chemical Theory and
  Computation} \textbf{2018}, \emph{14}, 1206--1215\relax
\mciteBstWouldAddEndPuncttrue
\mciteSetBstMidEndSepPunct{\mcitedefaultmidpunct}
{\mcitedefaultendpunct}{\mcitedefaultseppunct}\relax
\EndOfBibitem
\bibitem[Schneider \latin{et~al.}(2017)Schneider, Dai, Topper, Drechsel-Grau,
  and Tuckerman]{Schneider:2017}
Schneider,~E.; Dai,~L.; Topper,~R.~Q.; Drechsel-Grau,~C.; Tuckerman,~M.~E.
  Stochastic Neural Network Approach for Learning High-Dimensional Free Energy
  Surfaces. \emph{Phys. Rev. Lett.} \textbf{2017}, \emph{119}, 150601\relax
\mciteBstWouldAddEndPuncttrue
\mciteSetBstMidEndSepPunct{\mcitedefaultmidpunct}
{\mcitedefaultendpunct}{\mcitedefaultseppunct}\relax
\EndOfBibitem
\bibitem[Stecher \latin{et~al.}(2014)Stecher, Bernstein, and
  Csányi]{Stecher:2014}
Stecher,~T.; Bernstein,~N.; Csányi,~G. Free Energy Surface Reconstruction from
  Umbrella Samples Using Gaussian Process Regression. \emph{Journal of Chemical
  Theory and Computation} \textbf{2014}, \emph{10}, 4079--4097\relax
\mciteBstWouldAddEndPuncttrue
\mciteSetBstMidEndSepPunct{\mcitedefaultmidpunct}
{\mcitedefaultendpunct}{\mcitedefaultseppunct}\relax
\EndOfBibitem
\bibitem[Chen \latin{et~al.}(2015)Chen, Yu, and Tuckerman]{Chen:2015}
Chen,~M.; Yu,~T.-Q.; Tuckerman,~M. {Locating landmarks on high-dimensional free
  energy surfaces}. \emph{PNAS} \textbf{2015}, \emph{112}, 3235--3240\relax
\mciteBstWouldAddEndPuncttrue
\mciteSetBstMidEndSepPunct{\mcitedefaultmidpunct}
{\mcitedefaultendpunct}{\mcitedefaultseppunct}\relax
\EndOfBibitem
\bibitem[Zheng \latin{et~al.}(2013)Zheng, Rohrdanz, and Clementi]{Zheng:2013}
Zheng,~W.; Rohrdanz,~M.~A.; Clementi,~C. Rapid Exploration of Configuration
  Space with Diffusion-Map-Directed Molecular Dynamics. \emph{The Journal of
  Physical Chemistry B} \textbf{2013}, \emph{117}, 12769--12776\relax
\mciteBstWouldAddEndPuncttrue
\mciteSetBstMidEndSepPunct{\mcitedefaultmidpunct}
{\mcitedefaultendpunct}{\mcitedefaultseppunct}\relax
\EndOfBibitem
\bibitem[Chiavazzo \latin{et~al.}(2017)Chiavazzo, Covino, Coifman, Gear,
  Georgiou, Hummer, and Kevrekidis]{Chiavazzo:2017}
Chiavazzo,~E.; Covino,~R.; Coifman,~R.; Gear,~C.; Georgiou,~A.; Hummer,~G.;
  Kevrekidis,~I. {Intrinsic map dynamics exploration for uncharted effective
  free-energy landscapes}. \emph{PNAS} \textbf{2017}, \emph{114},
  E5494--E5503\relax
\mciteBstWouldAddEndPuncttrue
\mciteSetBstMidEndSepPunct{\mcitedefaultmidpunct}
{\mcitedefaultendpunct}{\mcitedefaultseppunct}\relax
\EndOfBibitem
\bibitem[Georgiou \latin{et~al.}(2017)Georgiou, Bello-Rivas, Gear, Wu,
  Chiavazzo, and Kevrekidis]{Georgiou:2017}
Georgiou,~A.~S.; Bello-Rivas,~J.~M.; Gear,~C.~W.; Wu,~H.-T.; Chiavazzo,~E.;
  Kevrekidis,~I.~G. An Exploration Algorithm for Stochastic Simulators Driven
  by Energy Gradients. \emph{Entropy} \textbf{2017}, \emph{19}, 294\relax
\mciteBstWouldAddEndPuncttrue
\mciteSetBstMidEndSepPunct{\mcitedefaultmidpunct}
{\mcitedefaultendpunct}{\mcitedefaultseppunct}\relax
\EndOfBibitem
\bibitem[Han \latin{et~al.}(2018)Han, Zhang, Car, and E]{Han:2018}
Han,~J.; Zhang,~L.; Car,~R.; E,~W. {Deep Potential: A General Representation of
  a Many-Body Potential Energy Surface}. \emph{Commun. Comput. Phys.}
  \textbf{2018}, \emph{23}, 629--639\relax
\mciteBstWouldAddEndPuncttrue
\mciteSetBstMidEndSepPunct{\mcitedefaultmidpunct}
{\mcitedefaultendpunct}{\mcitedefaultseppunct}\relax
\EndOfBibitem
\bibitem[Rupp \latin{et~al.}(2012)Rupp, Tkatchenko, M\"uller, and von
  Lilienfeld]{Rupp:2012}
Rupp,~M.; Tkatchenko,~A.; M\"uller,~K.-R.; von Lilienfeld,~O.~A. Fast and
  Accurate Modeling of Molecular Atomization Energies with Machine Learning.
  \emph{Phys. Rev. Lett.} \textbf{2012}, \emph{108}, 058301\relax
\mciteBstWouldAddEndPuncttrue
\mciteSetBstMidEndSepPunct{\mcitedefaultmidpunct}
{\mcitedefaultendpunct}{\mcitedefaultseppunct}\relax
\EndOfBibitem
\bibitem[Hansen \latin{et~al.}(2013)Hansen, Montavon, Biegler, Fazli, Rupp,
  Scheffler, von Lilienfeld, Tkatchenko, and Müller]{Hansen:2013}
Hansen,~K.; Montavon,~G.; Biegler,~F.; Fazli,~S.; Rupp,~M.; Scheffler,~M.; von
  Lilienfeld,~O.~A.; Tkatchenko,~A.; Müller,~K.-R. Assessment and Validation
  of Machine Learning Methods for Predicting Molecular Atomization Energies.
  \emph{Journal of Chemical Theory and Computation} \textbf{2013}, \emph{9},
  3404--3419\relax
\mciteBstWouldAddEndPuncttrue
\mciteSetBstMidEndSepPunct{\mcitedefaultmidpunct}
{\mcitedefaultendpunct}{\mcitedefaultseppunct}\relax
\EndOfBibitem
\bibitem[Hansen \latin{et~al.}(2015)Hansen, Biegler, Ramakrishnan, Pronobis,
  von Lilienfeld, M\"uller, and Tkatchenko]{Hansen:2015}
Hansen,~K.; Biegler,~F.; Ramakrishnan,~R.; Pronobis,~W.; von Lilienfeld,~O.~A.;
  M\"uller,~K.-R.; Tkatchenko,~A. Machine Learning Predictions of Molecular
  Properties: Accurate Many-Body Potentials and Nonlocality in Chemical Space.
  \emph{The Journal of Physical Chemistry Letters} \textbf{2015}, \emph{6},
  2326--2331, PMID: 26113956\relax
\mciteBstWouldAddEndPuncttrue
\mciteSetBstMidEndSepPunct{\mcitedefaultmidpunct}
{\mcitedefaultendpunct}{\mcitedefaultseppunct}\relax
\EndOfBibitem
\bibitem[Weininger \latin{et~al.}(1989)Weininger, Weininger, and
  Weininger]{Weininger:1989}
Weininger,~D.; Weininger,~A.; Weininger,~J.~L. SMILES. 2. Algorithm for
  generation of unique SMILES notation. \emph{Journal of Chemical Information
  and Computer Sciences} \textbf{1989}, \emph{29}, 97--101\relax
\mciteBstWouldAddEndPuncttrue
\mciteSetBstMidEndSepPunct{\mcitedefaultmidpunct}
{\mcitedefaultendpunct}{\mcitedefaultseppunct}\relax
\EndOfBibitem
\bibitem[Sanchez-Lengeling and Aspuru-Guzik(2018)Sanchez-Lengeling, and
  Aspuru-Guzik]{Sanchez-Lengeling:2018}
Sanchez-Lengeling,~B.; Aspuru-Guzik,~A. Inverse molecular design using machine
  learning: Generative models for matter engineering. \emph{Science}
  \textbf{2018}, \emph{361}, 360--365\relax
\mciteBstWouldAddEndPuncttrue
\mciteSetBstMidEndSepPunct{\mcitedefaultmidpunct}
{\mcitedefaultendpunct}{\mcitedefaultseppunct}\relax
\EndOfBibitem
\bibitem[Bart\'ok \latin{et~al.}(2010)Bart\'ok, Payne, Kondor, and
  Cs\'anyi]{Bartok:2010}
Bart\'ok,~A.~P.; Payne,~M.~C.; Kondor,~R.; Cs\'anyi,~G. Gaussian Approximation
  Potentials: The Accuracy of Quantum Mechanics, without the Electrons.
  \emph{Phys. Rev. Lett.} \textbf{2010}, \emph{104}, 136403\relax
\mciteBstWouldAddEndPuncttrue
\mciteSetBstMidEndSepPunct{\mcitedefaultmidpunct}
{\mcitedefaultendpunct}{\mcitedefaultseppunct}\relax
\EndOfBibitem
\bibitem[Rogers and Hahn(2010)Rogers, and Hahn]{Rogers:2010}
Rogers,~D.; Hahn,~M. Extended-Connectivity Fingerprints. \emph{Journal of
  Chemical Information and Modeling} \textbf{2010}, \emph{50}, 742--754, PMID:
  20426451\relax
\mciteBstWouldAddEndPuncttrue
\mciteSetBstMidEndSepPunct{\mcitedefaultmidpunct}
{\mcitedefaultendpunct}{\mcitedefaultseppunct}\relax
\EndOfBibitem
\bibitem[Bart\'ok \latin{et~al.}(2013)Bart\'ok, Kondor, and
  Cs\'anyi]{Bartok:2013}
Bart\'ok,~A.~P.; Kondor,~R.; Cs\'anyi,~G. On representing chemical
  environments. \emph{Phys. Rev. B} \textbf{2013}, \emph{87}, 184115\relax
\mciteBstWouldAddEndPuncttrue
\mciteSetBstMidEndSepPunct{\mcitedefaultmidpunct}
{\mcitedefaultendpunct}{\mcitedefaultseppunct}\relax
\EndOfBibitem
\bibitem[Duvenaud \latin{et~al.}(2015)Duvenaud, Maclaurin, Iparraguirre,
  Bombarell, Hirzel, Aspuru-Guzik, and Adams]{Duvenaud:2015}
Duvenaud,~D.~K.; Maclaurin,~D.; Iparraguirre,~J.; Bombarell,~R.; Hirzel,~T.;
  Aspuru-Guzik,~A.; Adams,~R.~P. In \emph{Advances in Neural Information
  Processing Systems 28}; Cortes,~C., Lawrence,~N.~D., Lee,~D.~D.,
  Sugiyama,~M., Garnett,~R., Eds.; Curran Associates, Inc., 2015; pp
  2224--2232\relax
\mciteBstWouldAddEndPuncttrue
\mciteSetBstMidEndSepPunct{\mcitedefaultmidpunct}
{\mcitedefaultendpunct}{\mcitedefaultseppunct}\relax
\EndOfBibitem
\bibitem[Pham \latin{et~al.}(2018)Pham, Nguyen, Nguyen, Kino, Miyake, and
  Dam]{Pham:2018}
Pham,~T.-L.; Nguyen,~N.-D.; Nguyen,~V.-D.; Kino,~H.; Miyake,~T.; Dam,~H.-C.
  Learning structure-property relationship in crystalline materials: A study of
  lanthanide–transition metal alloys. \emph{The Journal of Chemical Physics}
  \textbf{2018}, \emph{148}, 204106\relax
\mciteBstWouldAddEndPuncttrue
\mciteSetBstMidEndSepPunct{\mcitedefaultmidpunct}
{\mcitedefaultendpunct}{\mcitedefaultseppunct}\relax
\EndOfBibitem
\bibitem[Hirn \latin{et~al.}(2017)Hirn, Mallat, and Poilvert]{Hirn:2017}
Hirn,~M.; Mallat,~S.; Poilvert,~N. {Wavelet Scattering Regression of Quantum
  Chemical Energies}. \emph{Multiscale Model. Simul.} \textbf{2017}, \emph{15},
  827--863\relax
\mciteBstWouldAddEndPuncttrue
\mciteSetBstMidEndSepPunct{\mcitedefaultmidpunct}
{\mcitedefaultendpunct}{\mcitedefaultseppunct}\relax
\EndOfBibitem
\bibitem[Sidky and Whitmer(2018)Sidky, and Whitmer]{Sidky:2018}
Sidky,~H.; Whitmer,~J.~K. Learning free energy landscapes using artificial
  neural networks. \emph{The Journal of Chemical Physics} \textbf{2018},
  \emph{148}, 104111\relax
\mciteBstWouldAddEndPuncttrue
\mciteSetBstMidEndSepPunct{\mcitedefaultmidpunct}
{\mcitedefaultendpunct}{\mcitedefaultseppunct}\relax
\EndOfBibitem
\bibitem[Tang \latin{et~al.}(2018)Tang, Zhang, and Karniadakis]{Tang:2018}
Tang,~Y.-H.; Zhang,~D.; Karniadakis,~G.~E. An atomistic fingerprint algorithm
  for learning ab initio molecular force fields. \emph{The Journal of Chemical
  Physics} \textbf{2018}, \emph{148}, 034101\relax
\mciteBstWouldAddEndPuncttrue
\mciteSetBstMidEndSepPunct{\mcitedefaultmidpunct}
{\mcitedefaultendpunct}{\mcitedefaultseppunct}\relax
\EndOfBibitem
\bibitem[Behler and Parrinello(2007)Behler, and Parrinello]{Behler:2007}
Behler,~J.; Parrinello,~M. {Generalized Neural-Network Representation of
  High-Dimensional Potential-Energy Surfaces}. \emph{Physical Review Letters}
  \textbf{2007}, \emph{98}, 146401\relax
\mciteBstWouldAddEndPuncttrue
\mciteSetBstMidEndSepPunct{\mcitedefaultmidpunct}
{\mcitedefaultendpunct}{\mcitedefaultseppunct}\relax
\EndOfBibitem
\bibitem[Behler(2011)]{Behler:2011}
Behler,~J. Atom-centered symmetry functions for constructing high-dimensional
  neural network potentials. \emph{J. Chem. Phys.} \textbf{2011}, \emph{134},
  074106\relax
\mciteBstWouldAddEndPuncttrue
\mciteSetBstMidEndSepPunct{\mcitedefaultmidpunct}
{\mcitedefaultendpunct}{\mcitedefaultseppunct}\relax
\EndOfBibitem
\bibitem[Behler(2015)]{Behler:2015}
Behler,~J. {Constructing High-Dimensional Neural Network Potentials: A Tutorial
  Review}. \emph{Int. J. Quant. Chem.} \textbf{2015}, \emph{115},
  1032--1050\relax
\mciteBstWouldAddEndPuncttrue
\mciteSetBstMidEndSepPunct{\mcitedefaultmidpunct}
{\mcitedefaultendpunct}{\mcitedefaultseppunct}\relax
\EndOfBibitem
\bibitem[Behler(2017)]{Behler:2017}
Behler,~J. Neural network potential-energy surfaces for atomistic simulations.
  \emph{Chem. Modell.} \textbf{2017}, \emph{7}, 1--41\relax
\mciteBstWouldAddEndPuncttrue
\mciteSetBstMidEndSepPunct{\mcitedefaultmidpunct}
{\mcitedefaultendpunct}{\mcitedefaultseppunct}\relax
\EndOfBibitem
\bibitem[Jose \latin{et~al.}(2012)Jose, Artrith, and Behler]{Jose:2012}
Jose,~K. V.~J.; Artrith,~N.; Behler,~J. Construction of high-dimensional neural
  network potentials using environment-dependent atom pairs. \emph{The Journal
  of Chemical Physics} \textbf{2012}, \emph{136}, 194111\relax
\mciteBstWouldAddEndPuncttrue
\mciteSetBstMidEndSepPunct{\mcitedefaultmidpunct}
{\mcitedefaultendpunct}{\mcitedefaultseppunct}\relax
\EndOfBibitem
\bibitem[Cubuk \latin{et~al.}(2017)Cubuk, Malone, Onat, Waterland, and
  Kaxiras]{Cubuk:2017}
Cubuk,~E.; Malone,~B.; Onat,~B.; Waterland,~A.; Kaxiras,~E. {Representations in
  neural network based empirical potentials}. \emph{J. Chem. Phys.}
  \textbf{2017}, \emph{147}, 024104\relax
\mciteBstWouldAddEndPuncttrue
\mciteSetBstMidEndSepPunct{\mcitedefaultmidpunct}
{\mcitedefaultendpunct}{\mcitedefaultseppunct}\relax
\EndOfBibitem
\bibitem[Khaliullin \latin{et~al.}(2011)Khaliullin, Eshet, K{\"u}hne, Behler,
  and Parrinello]{Khaliullin:2011}
Khaliullin,~R.~Z.; Eshet,~H.; K{\"u}hne,~T.~D.; Behler,~J.; Parrinello,~M.
  Nucleation mechanism for the direct graphite-to-diamond phase transition.
  \emph{Nature Materials} \textbf{2011}, \emph{10}, 693\relax
\mciteBstWouldAddEndPuncttrue
\mciteSetBstMidEndSepPunct{\mcitedefaultmidpunct}
{\mcitedefaultendpunct}{\mcitedefaultseppunct}\relax
\EndOfBibitem
\bibitem[Behler \latin{et~al.}(2008)Behler, Marto\v{n}\'ak, Donadio, and
  Parrinello]{Behler:2008}
Behler,~J.; Marto\v{n}\'ak,~R.; Donadio,~D.; Parrinello,~M. Metadynamics
  Simulations of the High-Pressure Phases of Silicon Employing a
  High-Dimensional Neural Network Potential. \emph{Phys. Rev. Lett.}
  \textbf{2008}, \emph{100}, 185501\relax
\mciteBstWouldAddEndPuncttrue
\mciteSetBstMidEndSepPunct{\mcitedefaultmidpunct}
{\mcitedefaultendpunct}{\mcitedefaultseppunct}\relax
\EndOfBibitem
\bibitem[Artrith and Behler(2012)Artrith, and Behler]{Artrith:2012}
Artrith,~N.; Behler,~J. High-dimensional neural network potentials for metal
  surfaces: A prototype study for copper. \emph{Phys. Rev. B} \textbf{2012},
  \emph{85}, 045439\relax
\mciteBstWouldAddEndPuncttrue
\mciteSetBstMidEndSepPunct{\mcitedefaultmidpunct}
{\mcitedefaultendpunct}{\mcitedefaultseppunct}\relax
\EndOfBibitem
\bibitem[Artrith and Kolpak(2014)Artrith, and Kolpak]{Artrith:2014}
Artrith,~N.; Kolpak,~A.~M. Understanding the Composition and Activity of
  Electrocatalytic Nanoalloys in Aqueous Solvents: A Combination of DFT and
  Accurate Neural Network Potentials. \emph{Nano Letters} \textbf{2014},
  \emph{14}, 2670--2676, PMID: 24742028\relax
\mciteBstWouldAddEndPuncttrue
\mciteSetBstMidEndSepPunct{\mcitedefaultmidpunct}
{\mcitedefaultendpunct}{\mcitedefaultseppunct}\relax
\EndOfBibitem
\bibitem[Galvelis and Sugita(2017)Galvelis, and Sugita]{Galvelis:2017}
Galvelis,~R.; Sugita,~Y. Neural Network and Nearest Neighbor Algorithms for
  Enhancing Sampling of Molecular Dynamics. \emph{Journal of Chemical Theory
  and Computation} \textbf{2017}, \emph{13}, 2489--2500\relax
\mciteBstWouldAddEndPuncttrue
\mciteSetBstMidEndSepPunct{\mcitedefaultmidpunct}
{\mcitedefaultendpunct}{\mcitedefaultseppunct}\relax
\EndOfBibitem
\bibitem[Onat \latin{et~al.}(2018)Onat, Cubuk, Malone, and Kaxiras]{Onat:2018}
Onat,~B.; Cubuk,~E.~D.; Malone,~B.~D.; Kaxiras,~E. Implanted neural network
  potentials: Application to Li-Si alloys. \emph{Phys. Rev. B} \textbf{2018},
  \emph{97}, 094106\relax
\mciteBstWouldAddEndPuncttrue
\mciteSetBstMidEndSepPunct{\mcitedefaultmidpunct}
{\mcitedefaultendpunct}{\mcitedefaultseppunct}\relax
\EndOfBibitem
\bibitem[Smith \latin{et~al.}(2017)Smith, Isayev, and Roitberg]{Smith:2017}
Smith,~J.~S.; Isayev,~O.; Roitberg,~A.~E. {ANI-1: an extensible neural network
  potential with DFT accuracy at force field computational cost}. \emph{Chem.
  Sci.} \textbf{2017}, \emph{8}, 3192--3203\relax
\mciteBstWouldAddEndPuncttrue
\mciteSetBstMidEndSepPunct{\mcitedefaultmidpunct}
{\mcitedefaultendpunct}{\mcitedefaultseppunct}\relax
\EndOfBibitem
\bibitem[Smith \latin{et~al.}(2018)Smith, Nebgen, Lubbers, Isayev, and
  Roitberg]{Smith:2018}
Smith,~J.~S.; Nebgen,~B.; Lubbers,~N.; Isayev,~O.; Roitberg,~A.~E. Less is
  more: Sampling chemical space with active learning. \emph{The Journal of
  Chemical Physics} \textbf{2018}, \emph{148}, 241733\relax
\mciteBstWouldAddEndPuncttrue
\mciteSetBstMidEndSepPunct{\mcitedefaultmidpunct}
{\mcitedefaultendpunct}{\mcitedefaultseppunct}\relax
\EndOfBibitem
\bibitem[Chmiela \latin{et~al.}(2017)Chmiela, Tkatchenko, Sauceda, Poltavsky,
  Sch{\"{u}}tt, M{\"{u}}ller, Poltavsky, and Sch]{Chmiela:2017}
Chmiela,~S.; Tkatchenko,~A.; Sauceda,~H.~E.; Poltavsky,~I.;
  Sch{\"{u}}tt,~K.~T.; M{\"{u}}ller,~K.-r.; Poltavsky,~I.; Sch,~K.~T. {Machine
  learning of accurate energy-conserving molecular force fields}. \emph{Science
  Advances} \textbf{2017}, \emph{3}, e1603015\relax
\mciteBstWouldAddEndPuncttrue
\mciteSetBstMidEndSepPunct{\mcitedefaultmidpunct}
{\mcitedefaultendpunct}{\mcitedefaultseppunct}\relax
\EndOfBibitem
\bibitem[Christensen and von Lilienfeld(2020)Christensen, and von
  Lilienfeld]{Christensen:2020}
Christensen,~A.~S.; von Lilienfeld,~O.~A. On the role of gradients for machine
  learning of molecular energies and forces. \emph{Machine Learning: Science
  and Technology} \textbf{2020}, \emph{1}, 045018\relax
\mciteBstWouldAddEndPuncttrue
\mciteSetBstMidEndSepPunct{\mcitedefaultmidpunct}
{\mcitedefaultendpunct}{\mcitedefaultseppunct}\relax
\EndOfBibitem
\bibitem[Bartók and Csányi(2015)Bartók, and Csányi]{Bartok:2015}
Bartók,~A.~P.; Csányi,~G. Gaussian approximation potentials: A brief tutorial
  introduction. \emph{International Journal of Quantum Chemistry}
  \textbf{2015}, \emph{115}, 1051--1057\relax
\mciteBstWouldAddEndPuncttrue
\mciteSetBstMidEndSepPunct{\mcitedefaultmidpunct}
{\mcitedefaultendpunct}{\mcitedefaultseppunct}\relax
\EndOfBibitem
\bibitem[Huang \latin{et~al.}(2017)Huang, Shang, Zhang, and Liu]{Huang:2017}
Huang,~S.-D.; Shang,~C.; Zhang,~X.-J.; Liu,~Z.-P. Material discovery by
  combining stochastic surface walking global optimization with a neural
  network. \emph{Chem. Sci.} \textbf{2017}, \emph{8}, 6327--6337\relax
\mciteBstWouldAddEndPuncttrue
\mciteSetBstMidEndSepPunct{\mcitedefaultmidpunct}
{\mcitedefaultendpunct}{\mcitedefaultseppunct}\relax
\EndOfBibitem
\bibitem[Cooper \latin{et~al.}(2020)Cooper, Kästner, Urban, and
  Artrith]{Cooper:2020}
Cooper,~A.~M.; Kästner,~J.; Urban,~A.; Artrith,~N. Efficient training of ANN
  potentials by including atomic forces via Taylor expansion and application to
  water and a transition-metal oxide. \emph{npj Computational Materials}
  \textbf{2020}, \emph{6}, 54\relax
\mciteBstWouldAddEndPuncttrue
\mciteSetBstMidEndSepPunct{\mcitedefaultmidpunct}
{\mcitedefaultendpunct}{\mcitedefaultseppunct}\relax
\EndOfBibitem
\bibitem[Pukrittayakamee \latin{et~al.}(2009)Pukrittayakamee, Malshe, Hagan,
  Raff, Narulkar, Bukkapatnum, and Komanduri]{Pukrittayakamee:2009}
Pukrittayakamee,~A.; Malshe,~M.; Hagan,~M.; Raff,~L.~M.; Narulkar,~R.;
  Bukkapatnum,~S.; Komanduri,~R. Simultaneous fitting of a potential-energy
  surface and its corresponding force fields using feedforward neural networks.
  \emph{The Journal of Chemical Physics} \textbf{2009}, \emph{130},
  134101\relax
\mciteBstWouldAddEndPuncttrue
\mciteSetBstMidEndSepPunct{\mcitedefaultmidpunct}
{\mcitedefaultendpunct}{\mcitedefaultseppunct}\relax
\EndOfBibitem
\bibitem[Nguyen and Le(2012)Nguyen, and Le]{Nguyen:2012}
Nguyen,~H. T.~T.; Le,~H.~M. Modified Feed-Forward Neural Network Structures and
  Combined-Function-Derivative Approximations Incorporating Exchange Symmetry
  for Potential Energy Surface Fitting. \emph{The Journal of Physical Chemistry
  A} \textbf{2012}, \emph{116}, 4629--4638, PMID: 22548349\relax
\mciteBstWouldAddEndPuncttrue
\mciteSetBstMidEndSepPunct{\mcitedefaultmidpunct}
{\mcitedefaultendpunct}{\mcitedefaultseppunct}\relax
\EndOfBibitem
\bibitem[Nguyen-Truong and Le(2015)Nguyen-Truong, and Le]{Nguyen:2015}
Nguyen-Truong,~H.~T.; Le,~H.~M. An implementation of the Levenberg–Marquardt
  algorithm for simultaneous-energy-gradient fitting using two-layer
  feed-forward neural networks. \emph{Chemical Physics Letters} \textbf{2015},
  \emph{629}, 40--45\relax
\mciteBstWouldAddEndPuncttrue
\mciteSetBstMidEndSepPunct{\mcitedefaultmidpunct}
{\mcitedefaultendpunct}{\mcitedefaultseppunct}\relax
\EndOfBibitem
\bibitem[Nandi \latin{et~al.}(2019)Nandi, Qu, and Bowman]{Nandi:2019a}
Nandi,~A.; Qu,~C.; Bowman,~J.~M. Using Gradients in Permutationally Invariant
  Polynomial Potential Fitting: A Demonstration for CH4 Using as Few as 100
  Configurations. \emph{Journal of Chemical Theory and Computation}
  \textbf{2019}, \emph{15}, 2826--2835, PMID: 30896950\relax
\mciteBstWouldAddEndPuncttrue
\mciteSetBstMidEndSepPunct{\mcitedefaultmidpunct}
{\mcitedefaultendpunct}{\mcitedefaultseppunct}\relax
\EndOfBibitem
\bibitem[Conte \latin{et~al.}(2020)Conte, Qu, Houston, and Bowman]{Conte:2020}
Conte,~R.; Qu,~C.; Houston,~P.~L.; Bowman,~J.~M. Efficient Generation of
  Permutationally Invariant Potential Energy Surfaces for Large Molecules.
  \emph{Journal of Chemical Theory and Computation} \textbf{2020}, \emph{16},
  3264--3272, PMID: 32212729\relax
\mciteBstWouldAddEndPuncttrue
\mciteSetBstMidEndSepPunct{\mcitedefaultmidpunct}
{\mcitedefaultendpunct}{\mcitedefaultseppunct}\relax
\EndOfBibitem
\bibitem[Bowman \latin{et~al.}(2022)Bowman, Qu, Conte, Nandi, Houston, and
  Yu]{Bowman:2022}
Bowman,~J.~M.; Qu,~C.; Conte,~R.; Nandi,~A.; Houston,~P.~L.; Yu,~Q. The MD17
  datasets from the perspective of datasets for gas-phase “small” molecule
  potentials. \emph{J. Chem. Phys} \textbf{2022}, \emph{156}, 240901\relax
\mciteBstWouldAddEndPuncttrue
\mciteSetBstMidEndSepPunct{\mcitedefaultmidpunct}
{\mcitedefaultendpunct}{\mcitedefaultseppunct}\relax
\EndOfBibitem
\bibitem[Zaverkin \latin{et~al.}(2023)Zaverkin, Holzmüller, Bonfirraro, and
  Kästner]{Zaverkin:2023}
Zaverkin,~V.; Holzmüller,~D.; Bonfirraro,~L.; Kästner,~J. Transfer learning
  for chemically accurate interatomic neural network potentials. \emph{Phys.
  Chem. Chem. Phys.} \textbf{2023}, \emph{25}, 5383--5396\relax
\mciteBstWouldAddEndPuncttrue
\mciteSetBstMidEndSepPunct{\mcitedefaultmidpunct}
{\mcitedefaultendpunct}{\mcitedefaultseppunct}\relax
\EndOfBibitem
\bibitem[Singraber \latin{et~al.}(2019)Singraber, Morawietz, Behler, and
  Dellago]{Singraber:2019}
Singraber,~A.; Morawietz,~T.; Behler,~J.; Dellago,~C. Parallel Multistream
  Training of High-Dimensional Neural Network Potentials. \emph{Journal of
  Chemical Theory and Computation} \textbf{2019}, \emph{15}, 3075--3092\relax
\mciteBstWouldAddEndPuncttrue
\mciteSetBstMidEndSepPunct{\mcitedefaultmidpunct}
{\mcitedefaultendpunct}{\mcitedefaultseppunct}\relax
\EndOfBibitem
\bibitem[Gastegger \latin{et~al.}(2017)Gastegger, Behler, and
  Marquetand]{Gastegger:2017}
Gastegger,~M.; Behler,~J.; Marquetand,~P. Machine learning molecular dynamics
  for the simulation of infrared spectra. \emph{Chem. Sci.} \textbf{2017},
  \emph{8}, 6924--6935\relax
\mciteBstWouldAddEndPuncttrue
\mciteSetBstMidEndSepPunct{\mcitedefaultmidpunct}
{\mcitedefaultendpunct}{\mcitedefaultseppunct}\relax
\EndOfBibitem
\bibitem[Nandi \latin{et~al.}(2019)Nandi, Qu, and Bowman]{Nandi:2019b}
Nandi,~A.; Qu,~C.; Bowman,~J.~M. Full and fragmented permutationally invariant
  polynomial potential energy surfaces for trans and cis N-methyl acetamide and
  isomerization saddle points. \emph{J. Chem. Phys} \textbf{2019}, \emph{151},
  084306\relax
\mciteBstWouldAddEndPuncttrue
\mciteSetBstMidEndSepPunct{\mcitedefaultmidpunct}
{\mcitedefaultendpunct}{\mcitedefaultseppunct}\relax
\EndOfBibitem
\bibitem[Houston \latin{et~al.}(2020)Houston, Conte, Qu, and
  Bowman]{Houston:2020}
Houston,~P.; Conte,~R.; Qu,~C.; Bowman,~J.~M. Permutationally invariant
  polynomial potential energy surfaces for tropolone and H and D atom tunneling
  dynamics. \emph{J. Chem. Phys} \textbf{2020}, \emph{153}, 024107\relax
\mciteBstWouldAddEndPuncttrue
\mciteSetBstMidEndSepPunct{\mcitedefaultmidpunct}
{\mcitedefaultendpunct}{\mcitedefaultseppunct}\relax
\EndOfBibitem
\bibitem[Chen and Goldsmith(2020)Chen, and Goldsmith]{Chen:2020}
Chen,~X.; Goldsmith,~C.~F. Accelerating Variational Transition State Theory via
  Artificial Neural Networks. \emph{The Journal of Physical Chemistry A}
  \textbf{2020}, \emph{124}, 1038--1046\relax
\mciteBstWouldAddEndPuncttrue
\mciteSetBstMidEndSepPunct{\mcitedefaultmidpunct}
{\mcitedefaultendpunct}{\mcitedefaultseppunct}\relax
\EndOfBibitem
\bibitem[Unke \latin{et~al.}(2021)Unke, Chmiela, Gastegger, Schütt, Sauceda,
  and Müller]{Unke:2021b}
Unke,~O.~T.; Chmiela,~S.; Gastegger,~M.; Schütt,~K.~T.; Sauceda,~H.~E.;
  Müller,~K.-R. SpookyNet: Learning force fields with electronic degrees of
  freedom and nonlocal effects. \emph{Nature Communications} \textbf{2021},
  \emph{12}, 7273\relax
\mciteBstWouldAddEndPuncttrue
\mciteSetBstMidEndSepPunct{\mcitedefaultmidpunct}
{\mcitedefaultendpunct}{\mcitedefaultseppunct}\relax
\EndOfBibitem
\bibitem[Herr \latin{et~al.}(2018)Herr, Yao, McIntyre, Toth, and
  Parkhill]{Herr:2018}
Herr,~J.~E.; Yao,~K.; McIntyre,~R.; Toth,~D.~W.; Parkhill,~J. Metadynamics for
  training neural network model chemistries: A competitive assessment.
  \emph{The Journal of Chemical Physics} \textbf{2018}, \emph{148},
  241710\relax
\mciteBstWouldAddEndPuncttrue
\mciteSetBstMidEndSepPunct{\mcitedefaultmidpunct}
{\mcitedefaultendpunct}{\mcitedefaultseppunct}\relax
\EndOfBibitem
\bibitem[Najm and Yang(2021)Najm, and Yang]{Najm:2021}
Najm,~H.~N.; Yang,~Y. \emph{{AEVmod -- Atomic Environment Vector Module
  Documentation}}; 2021; Technical Report SAND2021--9473
  \url{https://doi.org/10.2172/1817835}\relax
\mciteBstWouldAddEndPuncttrue
\mciteSetBstMidEndSepPunct{\mcitedefaultmidpunct}
{\mcitedefaultendpunct}{\mcitedefaultseppunct}\relax
\EndOfBibitem
\bibitem[Paszke \latin{et~al.}(2019)Paszke, Gross, Massa, Lerer, Bradbury,
  Chanan, Killeen, Lin, Gimelshein, Antiga, Desmaison, Kopf, Yang, DeVito,
  Raison, Tejani, Chilamkurthy, Steiner, Fang, Bai, and Chintala]{pytorch:2019}
Paszke,~A. \latin{et~al.}  \emph{Advances in Neural Information Processing
  Systems 32}; Curran Associates, Inc., 2019; pp 8024--8035\relax
\mciteBstWouldAddEndPuncttrue
\mciteSetBstMidEndSepPunct{\mcitedefaultmidpunct}
{\mcitedefaultendpunct}{\mcitedefaultseppunct}\relax
\EndOfBibitem
\bibitem[Jakob \latin{et~al.}(2017)Jakob, Rhinelander, and Moldovan]{pybind11}
Jakob,~W.; Rhinelander,~J.; Moldovan,~D. \textsc{pybind11} -- Seamless
  operability between C++11 and Python. 2017;
  https://github.com/pybind/pybind11\relax
\mciteBstWouldAddEndPuncttrue
\mciteSetBstMidEndSepPunct{\mcitedefaultmidpunct}
{\mcitedefaultendpunct}{\mcitedefaultseppunct}\relax
\EndOfBibitem
\bibitem[Najm and Yang(2021)Najm, and Yang]{aevmod}
Najm,~H.~N.; Yang,~Y. \textsc{aevmod}. 2021;
  https://github.com/sandialabs/aevmod\relax
\mciteBstWouldAddEndPuncttrue
\mciteSetBstMidEndSepPunct{\mcitedefaultmidpunct}
{\mcitedefaultendpunct}{\mcitedefaultseppunct}\relax
\EndOfBibitem
\bibitem[Phipps(2022)]{Sacado}
Phipps,~E. \textsc{Sacado}. 2022; https://trilinos.github.io/sacado\relax
\mciteBstWouldAddEndPuncttrue
\mciteSetBstMidEndSepPunct{\mcitedefaultmidpunct}
{\mcitedefaultendpunct}{\mcitedefaultseppunct}\relax
\EndOfBibitem
\bibitem[qch()]{qchem:web}
https://www.q-chem.com\relax
\mciteBstWouldAddEndPuncttrue
\mciteSetBstMidEndSepPunct{\mcitedefaultmidpunct}
{\mcitedefaultendpunct}{\mcitedefaultseppunct}\relax
\EndOfBibitem
\bibitem[Kingma and Ba(2017)Kingma, and Ba]{kingma:2017}
Kingma,~D.~P.; Ba,~J. Adam: A Method for Stochastic Optimization. 2017;
  \url{https://doi.org/10.48550/arXiv.1412.6980}\relax
\mciteBstWouldAddEndPuncttrue
\mciteSetBstMidEndSepPunct{\mcitedefaultmidpunct}
{\mcitedefaultendpunct}{\mcitedefaultseppunct}\relax
\EndOfBibitem
\bibitem[pyt()]{pytorch:adam}
\url{https://pytorch.org/docs/stable/generated/torch.optim.Adam.html}\relax
\mciteBstWouldAddEndPuncttrue
\mciteSetBstMidEndSepPunct{\mcitedefaultmidpunct}
{\mcitedefaultendpunct}{\mcitedefaultseppunct}\relax
\EndOfBibitem
\end{mcitethebibliography}

\end{document}